\documentclass[
  journal=ancac3,
  manuscript=article
]{achemso}

\usepackage{graphicx}
\usepackage[T1]{fontenc}
\usepackage{bm}
\usepackage{siunitx}
\usepackage{booktabs}
\usepackage{soul}
\usepackage{amsmath}
\usepackage{amssymb}
\DeclareSIUnit\angstrom{\text {Å}}
\makeatletter
\renewcommand*\acs@contact@details{%
{ \sffamily \textsuperscript{*}Corresponding author. E-mail: \acs@email@list}%
\acs@number@list
}
\makeatother

\title{Valence-Selective Electrophoretic Transport in Neutral Carbon Nanotubes}

\author{Giuseppe Mattia Amato}
\affiliation[Northwestern MRSEC]
{Materials Research Science and Engineering Center, Northwestern University,
Evanston, Illinois 60208, United States}

\author{Felipe Jim\'enez-\'Angeles}
\affiliation[Northwestern MRSEC]
{Materials Research Science and Engineering Center, Northwestern University,
Evanston, Illinois 60208, United States}
\alsoaffiliation[Northwestern CCTSM]
{Center for Computation and Theory of Soft Materials, Northwestern University,
Evanston, Illinois 60208, United States}
\alsoaffiliation[Northwestern Materials Science]
{Department of Materials Science and Engineering, Northwestern University,
Evanston, Illinois 60208, United States}

\author{Monica Olvera de la Cruz}
\email{m-olvera@northwestern.edu}
\affiliation[Northwestern MRSEC]
{Materials Research Science and Engineering Center, Northwestern University,
Evanston, Illinois 60208, United States}
\alsoaffiliation[Northwestern CCTSM]
{Center for Computation and Theory of Soft Materials, Northwestern University,
Evanston, Illinois 60208, United States}
\alsoaffiliation[Northwestern Materials Science]
{Department of Materials Science and Engineering, Northwestern University,
Evanston, Illinois 60208, United States}
\alsoaffiliation[Northwestern Chemistry]
{Department of Chemistry, Northwestern University,
Evanston, Illinois 60208, United States}


\begin{document}

\begin{abstract}
Understanding how confinement alters the transport of charged molecules is central to the advancement of separation, energy, and sensing technologies. Polyamines are small, linear molecules with ionizable groups, making them an ideal model for investigating charge transport in confinement beyond monatomic ions. Using  all-atom molecular dynamics (MD) simulations, we investigate how molecular length, net charge, and charge placement affect the electrophoretic transport of polyamines through single-walled carbon nanotubes (CNTs). We find that nanometer-scale confinement reverses the mobility ordering observed in free solution---in a CNT of radius $\SI{0.68}{nm}$, spermine ($q=+4$) travels more slowly than ethylamine ($q=+1$), contrasting with the expected scaling of mobility with valence in bulk. Further, we find that mobility decreases faster than the inverse of valence, though molecular size and length vary alongside this trend. Comparisons of doubly protonated spermidine microspecies reveal the additional influence of charge placement at fixed net valence. Upon moving one proton of dicationic spermidine from the central to a terminal nitrogen, the mobility is increased by 42\%. We analyze hydration structure and solute--solvent association, identifying local-solvation differences between protonation topologies that accompany these distinct transport responses under confinement. These results suggest that neutral CNTs could support valence-selective nanofluidic transport, with additional sensitivity to molecular structure and protonation topology.
\end{abstract}

\maketitle

\begin{tocentry}
  \centering
  \includegraphics[width=3.25in,
  height=1.75in,
  keepaspectratio]{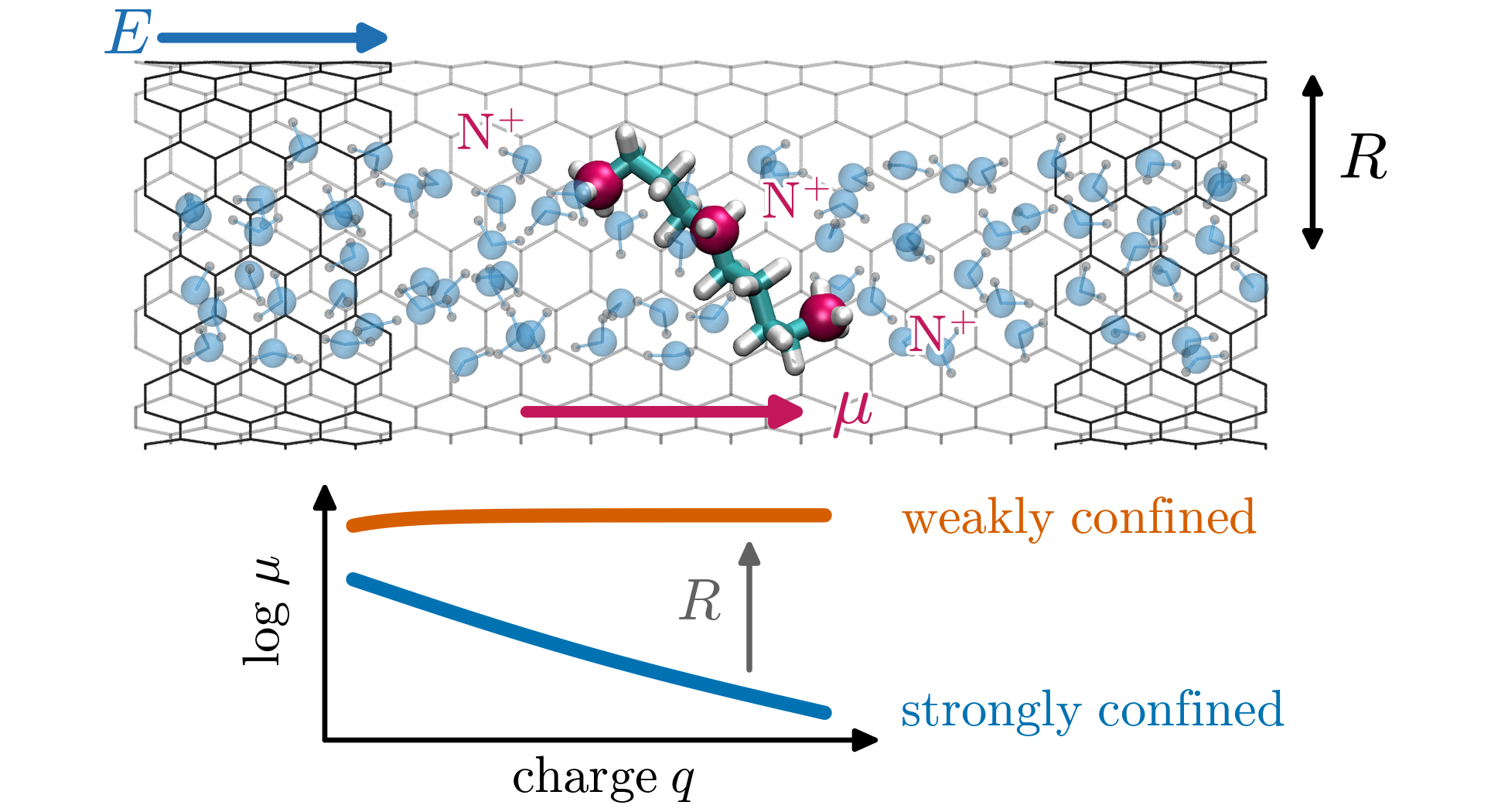}
\end{tocentry}

\section{Introduction}
\label{sec:intro}

Selective transport through nanoscale pores underpins both biological ion channels and synthetic membrane technologies. Pore geometry, surface charge, and chemistry of the local environment jointly control selectivity, which motivates model systems that distinguish their contributions.\cite{bocquet2010,sparreboom2009} An electric field can drive large charged molecules through nanopores, transiently modulating the ionic current, and thereby enabling detection of charged biopolymers. For example, alpha-hemolysin (a bacterial toxin) has allowed researchers to detect polynucleotides during translocation,\cite{kasianowicz1996} while MspA (an engineered pore) achieves single-nucleotide detection by coupling the translocation of DNA with a polymerase.\cite{manrao2012} The molecular selectivity achieved in these systems motivates understanding how individual pore and solute properties contribute to observed transport. 

Extending nanopore sensing to protein sequencing remains challenging, as amino acids differ in charge, size, conformation, and pore interactions. Recent experiments have demonstrated repeated nanopore reading of protein segments, and discrimination of individual amino acids in atomically thin pores.\cite{motone2024multipass,wang2023amino} Interpreting these signals, however, entails distinguishing the combined effects of molecular orientation, solvation, and solute--pore interactions. Simulations provide access to the molecular configurations associated with the measured response. Correspondingly, simulations of amino acids in a graphene nanopore further show that molecular orientation can strongly affect the current blockade,\cite{sur2025orientation} and simulations in carbon nanotubes (CNT) have identified excluded-volume and solute--wall interactions
as key factors in the selectivity of charged amino acids likewise.\cite{shen2025cnt} Beyond detecting molecules on the basis of current blockades, differences in their migration rates also are indicative of selective transport. A foundational question is then whether confinement alone can produce valence-selective transport. With cylindrical, graphitic walls and tunable radii, CNTs are useful candidates for addressing this question.

Electrophoretic mobility $\mu$ is the drift velocity divided by the applied electric field, $\mu=v/E$. For a particle with fixed hydrodynamic size, $\mu$ is directly proportional to the particle's charge.\cite{hunter1981} This expectation does not establish a universal ordering among different molecules, as external factors such as molecular size and conformation can also affect transport. For sufficiently long, uniformly charged polyelectrolytes in the free-draining regime, both the electrical driving force and friction scale approximately with chain length. This renders mobility approximately independent of polyelectrolyte length, and thereby hinders separation by free-solution electrophoresis.\cite{olvera1986electrophoresis,hoagland1999,stellwagen1997} In a gel, long chains stretch and cyclically compactify, again yielding length-independent mobility. Gel electrophoresis restores length dependence only for short chains.\cite{olvera1990pulsed}

Established mechanisms of nanoscale selectivity include surface electrostatics, direct pore--solute interactions, and dehydration or solvation barriers in finite pores.\cite{bocquet2010,zhang2026hydration} Wall chemistry influences both molecular attraction to the surface as well as the organization of surrounding water and its dielectric response. \cite{bonthuis2012permittivity} Hydrophilic or charged walls bind and reorder adjacent water and adsorb ions, whereas hydrophobic walls, such as those of CNTs, interact weakly with water, supporting low-friction fast transport.\cite{bocquet2010,secchi2016}

Molecular simulations have shown that electroosmotic flow driven by counterions can screen the effective charge of DNA in a nanopore\cite{luan2008} whereas the transverse confinement affects the polyelectrolyte's mobility.\cite{shendruk2015} 
More generally, strong confinement challenges assumptions underlying the  transport in bulk.\cite{bocquet2010,sparreboom2009} Namely, experimental measurements of water transport in CNTs reveal velocities far exceeding the no-slip continuum predictions, \cite{hummer2001,holt2006,majumder2005} and a strongly radius-dependent surface slippage.\cite{secchi2016} Simulations of conductive nanochannels show that surface polarization can modify conductivity and ionic correlations, as well as yield nonlinear electrical responses.\cite{dossantos2023polarizable,jimenez2023polarization} Recent simulations of electrolytes confined between graphene surfaces frame self-diffusion of ions and ion correlations as contributing factors in confined transport.\cite{fong2025} These observations motivate examining whether the charge--mobility relation can be altered under neutral confinement, and which molecular properties accompany that response.

We consider the transport of polyamines through single-walled CNTs. Polyamines play key roles in cell proliferation.\cite{tabor1984,igarashi2010} Notably, polyamines are also dysregulated in many cancers.\cite{pegg2016,casero2018} Polyamines are associated with the compaction of DNA, which is a process crucial to DNA packing in viruses.\cite{raspaud1998} The four molecules differ in size and structure as well as valence, preventing their differences in transport from being attributed uniquely to the direct contribution of charge. We therefore also examine different spermidine (a polyamine with maximum charge $q=+3$) protonation states on the same molecular backbone, including dicationic microspecies $\text{1,5-Spd}^{2+}$ and $\text{1,10-Spd}^{2+}$. The dicationic $\text{1,5-Spd}^{2+}$ has charged amine groups at the 1st and 5th positions, whereas the  $\text{1,10-Spd}^{2+}$ species has charged groups at the 1st and 10th positions. We refer to protonation states differing in net charge as macrospecies, and to different protonation states with identical net charge as microspecies. This serves to separate molecular identity from the combined changes in valence, ion counts, and protonation topology.

We first characterize the dependence of polyamine mobility on the degree of confinement and the applied field. We then examine transport differences across spermidine protonation states. Finally, we analyze hydration structure and solute--solvent association to identify the features accompanying these observed transport responses. These comparisons address valence-selective transport in neutral CNTs and the sensitivity of confined motion to protonation topology.

\section{Model system}
\label{sec:model}

\begin{figure}[tbp]
    \centering
    \includegraphics[width=0.5\linewidth]{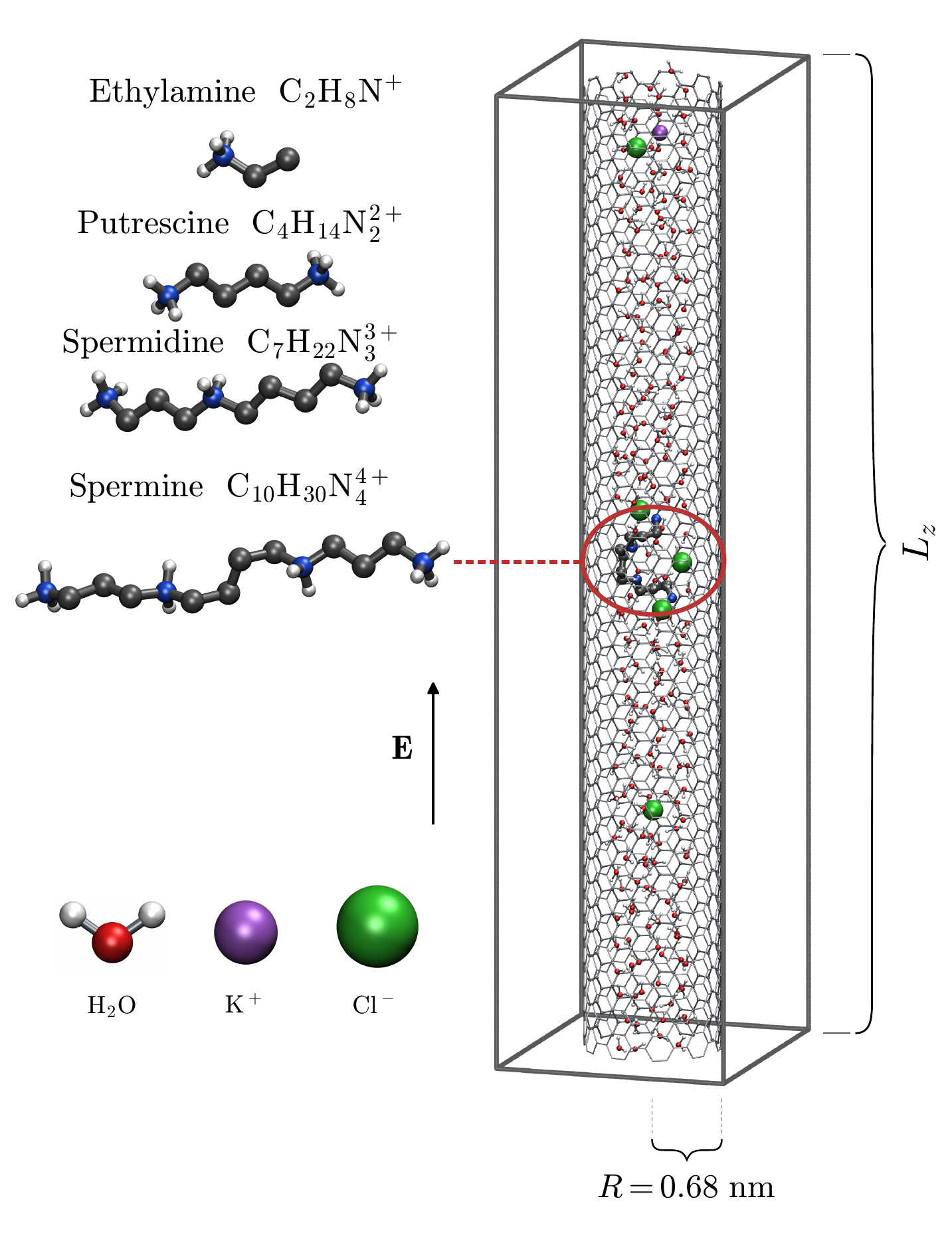}
    \caption{System setup, CNT geometry, and fully protonated polyamines. Hydrogen atoms bonded to carbon are omitted for clarity. The schematic identifies the system's cell length and tube radius; individual system sizes and compositions are presented in the SI.}
    \label{fig:setup}
\end{figure}

We employ all-atom nonequilibrium molecular dynamics (NEMD) to study the transport of polyamine species within a neutral, axially periodic carbon nanotube---an effectively infinite cylinder---with explicit water (see Figure \ref{fig:setup}). The uncharged wall removes the potential effects of imposed surface charge while the periodic CNT excludes tube-entry kinetics. We investigate the transport of ethylamine ($q=+1$), putrescine ($q=+2$), spermidine ($q=+3$), and spermine ($q=+4$) in single-walled CNTs considering four radii ($R=0.68$, $1.02$, $1.49$, and \SI{1.97}{\nano\meter}) and a range of applied electric fields. Our model retains both steric and dispersion interactions, providing a controlled setting to investigate how confinement influences valence-dependent transport within the pore. 

We simulate armchair $(10,10)$, $(15,15)$, $(22,22)$, and $(29,29)$ single-walled CNTs with geometric radii $R = 0.68$, $1.02$, $1.49$, and $\SI{1.97}{\nano\meter}$, respectively (Figure~\ref{fig:setup}). These radii describe the physical carbon cylinder, but not the radius which is accessible by solvent directly. Each cell contains one polyamine, explicit water, one to three added KCl pairs, and additional chloride ions required to maintain electroneutrality---all contained within the CNT. We vary the axial cell length and KCl count to obtain an approximately \SI{50}{mM} salt concentration; the Supporting Information specifies individual system compositions. The simulation details are described in the Methods section.

\section{Results and Discussion}
\label{sec:results}
\subsection{Valence-selective transport under confinement}

\begin{figure*}
  \centering
  \includegraphics[width=\columnwidth]{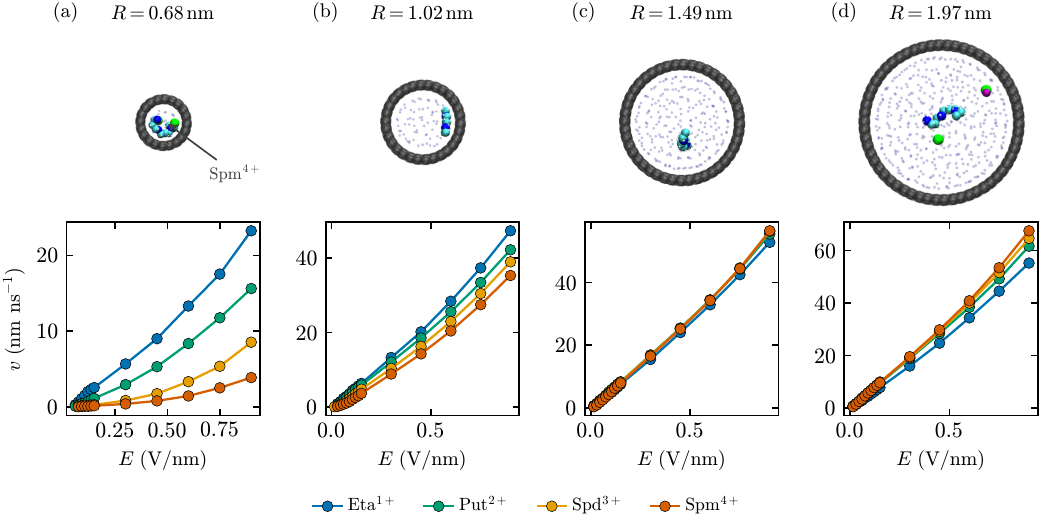}
  \caption{Drift velocity $v$ versus applied field $E$ for the four polyamines in CNTs of radius (a) $R = \SI{0.68}{\nano\meter}$, (b) $R = \SI{1.02}{\nano\meter}$, (c) $R = \SI{1.49}{\nano\meter}$, and (d) $R = \SI{1.97}{\nano\meter}$. In the narrowest CNT, the charge--velocity ordering is inverted; i.e. ethylamine ($q=+1$) migrates the fastest and spermine ($q=+4$) the slowest. This inversion weakens at $R = \SI{1.02}{\nano\meter}$; at $R = \SI{1.49}{\nano\meter}$ and $R = \SI{1.97}{\nano\meter}$, ethylamine becomes the slowest species. The transport response is markedly nonlinear in $E$ in the narrowest CNT. The top row shows simulation snapshots of $\mathrm{Spm}^{4+}$ in each CNT rendered with a uniform length scale. Colors identify the polyamines as per the key. Error bars are standard errors across three replicate simulations at $E=\SI{0.135}{\volt\per\nano\meter}$; otherwise, errors represent the standard error taken from a five-block division of the single trajectories. These errors are smaller than the plotted symbols.}
  \label{fig:velocity}
\end{figure*}

We first examine the drift velocities of the four polyamines as a function of the applied electric field $E$ across the four CNT radii. At any given electric field in bulk, the amines' velocities are determined by their charge and follow the order $v_{{4+}}$ > $v_{{3+}}$> $v_{{2+}}$ > $v_{{1+}}$; $v_{{1+}}$, $v_{{2+}}$, $v_{{3+}}$, and $v_{{4+}}$ represent the velocity of ethylamine, putrescine, spermidine, and spermine, respectively. We find that in the narrowest tube ($R = \SI{0.68}{\nano\meter}$, Figure~\ref{fig:velocity}a), the velocity--charge ordering is inverted relative to bulk; across each electric field strength, ethylamine migrates fastest, and spermine has the slowest velocity. Notably, not only is the order inverted, but the velocity difference between the amines is enhanced, which suggests a more efficient amine separation in confinement. For example, at $E = \SI{0.135}{\volt\per\nano\meter}$, ethylamine drifts at $v = 2.318 \pm \SI{0.030}{\nano\meter/ns}$ while spermine achieves $v = 0.143 \pm \SI{0.008}{\nano\meter/ns}$, despite carrying four times the charge of ethylamine. The corresponding  mobilities of the amines are
$\mu_{\mathrm{Eta}} = 17.17 \pm 0.22$ and $\mu_\text{Spm} = 1.06 \pm \SI{0.06}{\nano\meter^2/(V \cdot ns)}$, respectively, with uncertainties taken from triplicate simulation trials. This inverted ordering persists across simulated fields through to $\SI{0.900}{\volt\per\nano\meter}$. The simulation of fields within this high-field range was executed to probe the scaling of the observed inversion. Interestingly, the drift velocities follow a non-linear trend as a function of the applied external field; the field-dependence of $\mu=v/E$ reflects the effect of confinement. In the narrowest CNT, chloride associates more strongly with the multivalent amines (Section~\ref{sec:nonlinear})---the number of chlorides in contact exceeds uniform mixing by 14 times for $\mathrm{Spm}^{4+}$, whereas the corresponding factor for $\mathrm{Eta}^{1+}$ is 1.4. In free solution, the increase in association ranges from 1.2--1.8-fold for the polyamines. Ion association and correlations of this kind are strengthened by nanoconfinement and have been shown to give rise to nonlinear transport.\cite{dossantos2023polarizable,jimenez2023polarization} Association does not, however, generate the inversion---in configurations with no chloride contact, the inverted ordering persists at $E\geq\SI{0.450}{\volt\per\nano\meter}$ (Supporting Information, Section~S7).

Water velocities remain consistently small relative to the amine drift ($|v_{\rm water}/v_{\rm amine}|<0.08$, Figure~S2), which excludes solvent advection as the explanation for the amine ordering. Chloride drifts oppositely to the cationic polyamines---transport is likewise not a result of chloride co-advection. Neither observation excludes local hydrodynamic coupling or solute--counterion interactions.

In free solution, mobilities of ethylamine and spermine are $\mu = 50.7$ and $\SI{67.1}{\nano\meter^2/(V \cdot ns)}$, respectively, increasing monotonically with valence. We therefore find confinement in the neutral $R=\SI{0.68}{\nano\meter}$ CNT to transform a 1.3-fold increase in mobility from ethylamine to spermine (in free solution) to an approximately 16-fold decrease across the charge series. Figures~\ref{fig:velocity}b--d show the corresponding velocity $v$ as a function of the applied electric field $E$ in nanotubes of radius $R=1.02$, 1.49, and \SI{1.97}{\nano\meter}. The velocity trend inversion diminishes as confinement is relaxed---at $R=\SI{1.02}{\nano\meter}$, the ordering remains inverted but is less pronounced; $\mathrm{Spm}^{4+}$ migrates at roughly half the velocity of $\mathrm{Eta}^{1+}$. At $R=\SI{1.49}{\nano\meter}$ and $R=\SI{1.97}{\nano\meter}$, ethylamine becomes the slowest species above $E\approx\SI{0.1}{\volt\per\nano\meter}$, approaching the ordering in free solution. The crossover from inverted to standard ordering therefore occurs between $R=1.02$ and $R=\SI{1.49}{\nano\meter}$. Mobilities for each system at each applied field strength are reported in Figure~\ref{fig:nonlinear}a and Figure~S1.

\subsection{Confinement-dependent mobility and transport ordering}

\begin{figure*}
  \centering
  \includegraphics[width=\columnwidth]{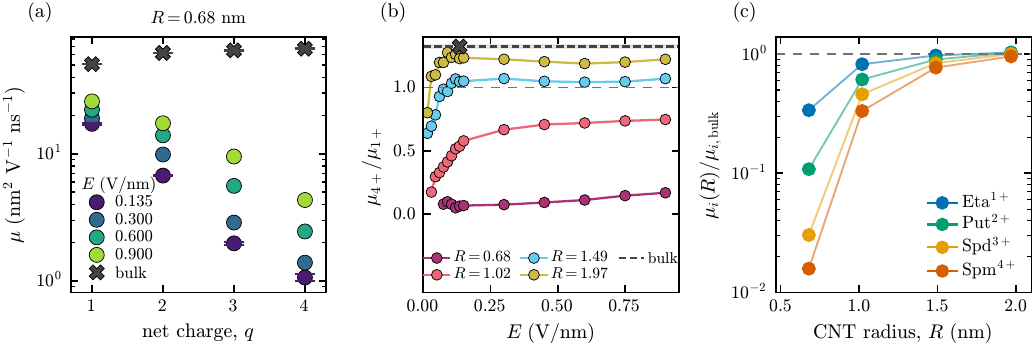}
  \caption{Confinement and field effects on valence selectivity. (a) Electrophoretic mobility $\mu$ as a function of polyamine charge in the $R=\SI{0.68}{\nano\meter}$ CNT at four field strengths, increasing from dark to light color. Free-solution mobilities at $E=\SI{0.135}{\volt\per\nano\meter}$ are depicted by crosses and are taken as the limit in the absence of confinement; these mobilities increase by at most 6\% at \SI{0.300}{\volt\per\nano\meter} (Supporting Information). Mobility decreases monotonically with charge under each field strength displayed. The spacing between the series of electric fields demonstrates the confinement-induced field dependence of $\mu$. (b) Spermine-to-ethylamine mobility ratio $\mu_{4+}/\mu_{1+}$ as a function of field strength for each CNT radius. The dashed line marks the ratio in free solution. Values of $\mu_{4+}/\mu_{1+} < 1$ denote an inversion in ordering. (c) Mobility of each species $i$ at radius $R$ and $E=\SI{0.135}{\volt\per\nano\meter}$ relative to the mobility of species $i$ in free solution. Unity denotes the corresponding mobility in free solution. The spread of mobility among species is the largest in the narrowest CNT.}
  \label{fig:resistance}
\end{figure*}

To evaluate the nanotubes' selectivity we investigate the amines' mobility $\mu=v/E$. 
Figure~\ref{fig:resistance}a compares electrophoretic mobility with charge in the narrowest CNT ($R=0.68$ nm). We observe that at each electric field strength, the amine's mobility follow an opposing trend from that in the bulk (see Figure \ref{fig:resistance}a). At $E=\SI{0.135}{\volt\per\nano\meter}$, with ethylamine as the reference, we observe $\mu(q)/\mu(q{=}1)$ of $0.39$, $0.11$, and $0.06$ for $q=+2$, $+3$, and $+4$, respectively. The electrophoretic mobility decreases more than proportionally with the inverse of amine charge ($1/q$).

At stronger electric fields the mobility continues to decrease with charge, but the relative differences among the amine mobilities diminish---as $E$ rises from $E=\SI{0.135}{\volt\per\nano\meter}$ to $E=\SI{0.900}{\volt\per\nano\meter}$, $\mu_{4+}/\mu_{1+}$ rises from 0.06 to 0.17. The absolute difference in mobility $\mu_{4+}-\mu_{1+}$ is enlarged from $-16.1$ to $-21.5$~\si{\nano\meter^2/(V.ns)}.

We quantify the nanotubes' selectivity between spermine and ethylamine by evaluating $\mu_{4+}/\mu_{1+}$ in nanotubes of different radii; the closer $\mu_{4+}/\mu_{1+}$ is to unity, the less selective the pore is. The field dependence of the mobility-charge relation $\mu_{4+}/\mu_{1+}$ is depicted in Figure~\ref{fig:resistance}b. We find that beyond the low-field regime ($E\geq\SI{0.300}{V/nm}$), the  mobility of $\mathrm{Spm}^{4+}$ increases roughly proportionally to the mobility increase of $\mathrm{Eta}^{1+}$, so their quotient plateaus for $R\geq\SI{1.02}{\nano\meter}$. The magnitude of this plateau scales with CNT radius, approaching the free-solution limit as confinement is relaxed---it remains between 0.06 and 0.17 at $R=\SI{0.68}{\nano\meter}$, continuing to rise with increased field only for the narrowest CNT, stays below unity at $R=\SI{1.02}{\nano\meter}$ (0.67--0.75), and slightly exceeds unity at $R = 1.49$ and \SI{1.97}{\nano\meter} (1.04--1.07 and 1.19--1.22, respectively). Thus, the CNT radius determines the mobility ordering while the field strength modulates its overall magnitude. All the investigated nanotubes have some degree of selectivity ($\mu_{4+}/\mu_{1+}\neq1$), which, for $R\leq\SI{1.49}{\nano\meter}$, is enhanced in the low-field regime.

Figure~\ref{fig:resistance}c illustrates the mobility deviation from the bulk value, by comparing the mobility of an amine in a particular CNT radius to its mobility in bulk. The deviation is not uniform. In the narrowest CNT at $E=\SI{0.135}{\volt\per\nano\meter}$, mobilities deviate from their free-solution magnitudes by factors of $2.95$, $9.23$, $33.1$, and $63.2$ for $\mathrm{Eta}^{1+}$ through $\mathrm{Spm}^{4+}$, respectively. The separation thereby occurs differentially rather than an overall diminishment in mobility. The spread between the amines is largely closed by $R = \SI{1.49}{\nano\meter}$.

\subsection{Modulation of transport behavior at fixed valence through alteration of charge placement}

\begin{figure*}
  \centering
  \includegraphics[width=\columnwidth]{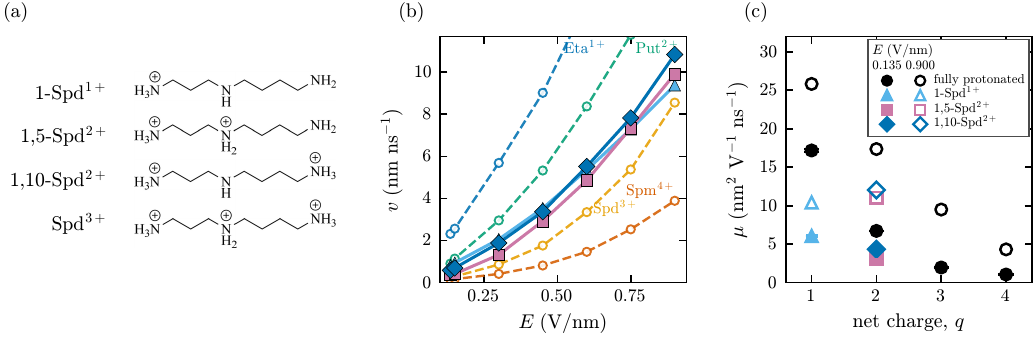}
  \caption{Transport properties of spermidine macrospecies at $R=\SI{0.68}{\nano\meter}$. (a) Structures of the simulated spermidine protonation topologies. (b) Drift velocity versus applied field for the charged spermidine variants in the $R=\SI{0.68}{\nano\meter}$ CNT (solid lines, filled symbols) compared with parent-amine velocities (dashed lines, open circles). (c) Mobility as a function of net charge for the parent amines (black circles) and the spermidine variants (colored symbols). The filled data points indicate $E=\SI{0.135}{\volt\per\nano\meter}$ and the open circles $E=\SI{0.900}{\volt\per\nano\meter}$. At low field, the mobilities of the spermidine variants decrease upon increasing net charge whereas with high field, the velocities converge and are no longer ordered by charge: both dicationic microspecies exceed $1\text{-}\mathrm{Spd}^{1+}$ and $\mathrm{Spd}^{3+}$ remains the least mobile. Increasing the strength of the electric field causes the mobility to increase in a charge-dependent manner. The distally protonated $q=+2$ topology remains distinct from the adjacently protonated microspecies.}
  \label{fig:protonation}
\end{figure*}

The four polyamines differ in charge, molecular size and weight. Therefore, the mobility trend cannot be attributed uniquely to the charge. To evaluate the roles of the charge, charge distribution, and molecular size, we examine three spermidine variations denoted as 1-Spd$^{1+}$, 1,5-Spd$^{2+}$, and 1,10-Spd$^{2+}$ (see Figure~\ref{fig:protonation}a). Increasing protonation requires additional chloride counterions and changes the protonation topology. The two dicationic microspecies, sharing the same molecular backbone and system composition (identical water, K$^+$, Cl$^-$ counts), function to investigate the charge placement effect.

Figure \ref{fig:protonation}b shows the velocities as a function of the applied electric field of the four parent amines and the spermidine variants in a CNT of radius \SI{0.68}{\nano\meter}. With $E\geq\SI{0.120}{\volt\per\nano\meter}$, the shorter but identically charged molecules (Eta$^{1+}$ and Put$^{2+}$) move faster than the $1\text{-}\mathrm{Spd}^{1+}$ and $\mathrm{Spd}^{2+}$ variants. Interestingly, the velocities of the variants with one or two deprotonated sites ($1\text{-}\mathrm{Spd}^{1+}$ and both $\mathrm{Spd}^{2+}$ microspecies) migrate with velocity that is clearly higher than that of the fully-protonated parent species (Spd$^{3+}$), converging at high field (\SI{0.900}{\volt\per\nano\meter}).  

The charge variants follow the same decrease in mobility with charge as the fully protonated polyamines in the \SI{0.68}{\nano\meter}-CNT. At $E = \SI{0.135}{\volt\per\nano\meter}$, the mobility decreases monotonically, from $\SI{6.06}{\nano\meter^2/(V.ns)}$ for $1\text{-}\mathrm{Spd}^{1+}$ to $4.33$ and $\SI{3.05}{\nano\meter^2/(V.ns)}$ for $1{,}10\text{-}\mathrm{Spd}^{2+}$ and $1{,}5\text{-}\mathrm{Spd}^{2+}$, respectively, and
$\SI{1.97}{\nano\meter^2/(V.ns)}$ for $\mathrm{Spd}^{3+}$ (Figure~\ref{fig:protonation}c). Comparisons between the spermidine variants and the parent amines with the same charge reveal that the transport response contains information beyond net valence. $1\text{-}\mathrm{Spd}^{1+}$ is less mobile than $\mathrm{Eta}^{1+}$, while $\mathrm{Put}^{2+}$ and both $\mathrm{Spd}^{2+}$ topologies exhibit distinct mobilities despite having equal charge. Therefore, in the narrowest CNT, transport is sensitive to charge state and molecular structure as well as the placement of charge. 

The two matched-charge microspecies of spermidine decouple net valence from charge distribution.
Across the simulated field strengths for $R = \SI{0.68}{\nano\meter}$ (excluding unresolved values $E<\SI{0.075}{\volt\per\nano\meter}$), the separated-site $1,10\text{-}{\mathrm{Spd}}^{2+}$ protonated
spermidine migrates $7-71\%$ faster than the adjacent-site $1,5\text{-}{\mathrm{Spd}}^{2+}$. At $E = \SI{0.135}{\volt\per\nano\meter}$, $\frac{v(1,10\text{-}\mathrm{Spd}^{2+})}{v(1,5\text{-}\mathrm{Spd}^{2+})}=1.42\pm0.05$. As both microspecies carry identical net charge and backbone, this demonstrates that relocating protonation sites (changing the charge distribution) alters the electrophoretic transport. The velocity response for each spermidine protonation state is reported in the Supporting Information (Figure~S1e).

All these phenomena highlight the roles of molecular size, net charge, and charge placement in the molecules' transport. Next, we examine the molecular hydration to elucidate the underlying mechanisms.

\subsection{Field-dependent counterion association and nonlinear transport}
\label{sec:nonlinear}

\begin{figure*}
    \centering
    \includegraphics[width=\columnwidth]{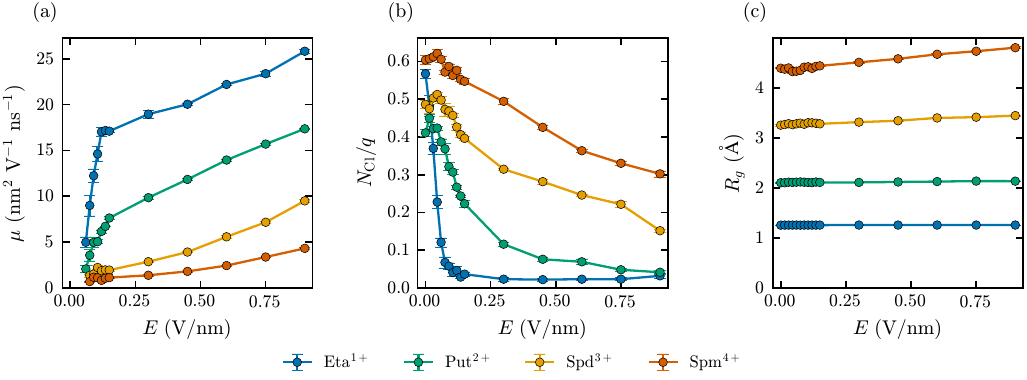}
    \caption{Polyamine transport response in confinement. (a) Electrophoretic mobility $\mu = v/E$, (b) coordination number
with chloride ions (per amine charge, $N_{\mathrm{Cl}^-}/q$), and (c) radius of gyration $R_g$ of polyamines as a function of the applied
external field, $E$. The amines are confined in a nanotube of \SI{0.68}{\nano\meter} in radius.}
\label{fig:nonlinear}
\end{figure*}

Beyond separation, the transport response of charged molecules in confinement is important for energy, information, and memory processes.\cite{robin2021} Figure~\ref{fig:nonlinear}a shows the polyamines’ electrophoretic mobility ($\mu=v/E$) as a function of the applied electric field in a \SI{0.68}{\nano\meter} CNT. The electrophoretic mobility trends imply a non-linear dependence of the drift velocity on the electric field, which differs for each amine. For $\mathrm{Eta}^{1+}$, the mobility is nearly linear but shows distinct slopes in two regimes ($E<\SI{0.150}{\volt\per\nano\meter}$ and $E>\SI{0.150}{\volt\per\nano\meter}$), whereas for $\mathrm{Put}^{2+}$, $\mathrm{Spd}^{3+}$, and $\mathrm{Spm}^{4+}$, the mobility is close to linear below $E\approx\SI{0.1}{\volt\per\nano\meter}$ and non-linear for stronger electric fields. The mobility behavior relates to polyamine association--dissociation with their $\mathrm{Cl}^-$ counterions. Figure~\ref{fig:nonlinear}b shows the average polyamine coordination number with $\mathrm{Cl}^-$ ions per number of charges, $q$ (counterion coordination is often termed ion condensation or ion association along the chain in the polyelectrolyte literature (in dilute solutions),\cite{gonzalezmozuelos1995,stevens1995}). At zero electric field, all polyamines show a high coordination number with $\mathrm{Cl}^-$ ions ($N_{\mathrm{Cl}^-}/q\sim0.5$), implying that $\mathrm{Cl}^-$ ions nearly screen their charge. For $\mathrm{Eta}^{1+}$, the coordination number drops sharply below 0.1 at electric fields larger than \SI{0.075}{\volt\per\nano\meter}, implying that the field induces $\mathrm{Cl}^-$ dissociation from protonated amine groups ($-\mathrm{NH}_3^{+}$). This dissociation explains the increase in $\mathrm{Eta}^{1+}$ mobility with the field and the two mobility regimes. $\mathrm{Put}^{2+}$, $\mathrm{Spd}^{3+}$, and $\mathrm{Spm}^{4+}$ maintain a coordination number above 0.2 up to $E=\SI{0.15}{\volt\per\nano\meter}$, and dissociate more gradually with increased field. Related phenomena have been observed in the ionic conductivity in polyelectrolyte hydrogels.\cite{li2016} There, the non-linear effects were also related to the change in ion association in strong fields. Our analysis shows that the electric field has little effect on the polyamines’ conformation, as demonstrated by the small relative change (at most 9\%) in their radius of gyration across the range of electric fields (see Figure~\ref{fig:nonlinear}c).

In Onsager’s theory of transport of weak electrolytes, deviations from the linear
response arise from the dissociation constant dependence on an external electric field.\cite{onsager1934} Monoatomic ions may exhibit non-linear
responses by forming pairs and elongated ``polyelectrolyte Wien'' chains in planar
confinement.\cite{robin2021} We show that non-linear mobility correlates with the field-dependent dissociation of $\mathrm{Cl}^-$ ions from protonated amine groups ($-\mathrm{NH}_3^{+}$), and that confinement and molecular length tune this dissociation. Our results suggest a strategy to separate small, charged molecules, which we speculate could be extended to separate longer polyelectrolyte chains.

\subsection{Charge topology redistributes hydration under confinement}

\begin{figure*}
  \centering
  \includegraphics[width=\columnwidth]{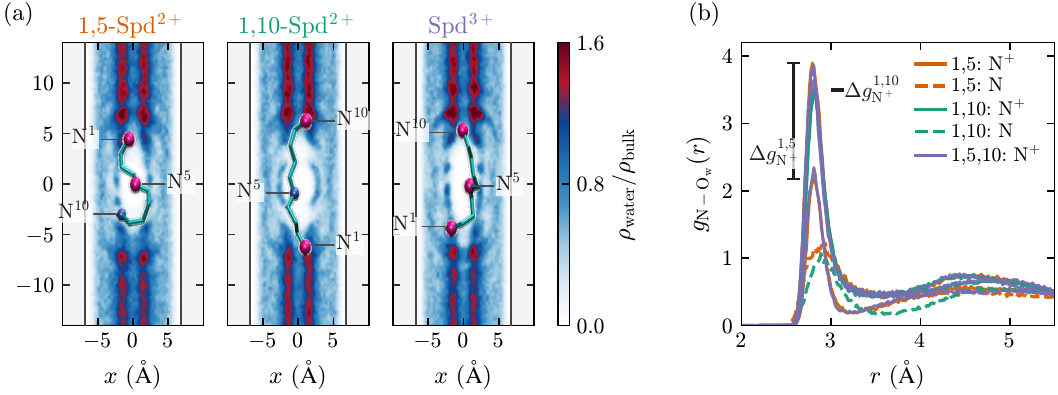}
  \caption{Spatial arrangement of charge alters confined hydration in spermidine protonation states. (a) Amine-COM-centered water-density maps at $E=\SI{0.135}{\volt\per\nano\meter}$ for $1{,}5\text{-}\mathrm{Spd}^{2+}$, $1{,}10\text{-}\mathrm{Spd}^{2+}$, and $\mathrm{Spd}^{3+}$, including representative conformations of the amines obtained directly from simulation trajectories. (b) Nitrogen--water-oxygen radial density ratios for the three nitrogen sites of each species. Orange denotes $1{,}5\text{-}\mathrm{Spd}^{2+}$, teal $1{,}10\text{-}\mathrm{Spd}^{2+}$, and purple $\mathrm{Spd}^{3+}$. Solid curves denote protonated nitrogen while dashed curves denote neutral sites. Both panels are averages over independent triplicate trials.}
  \label{fig:hydration}
\end{figure*}

The spermidine variants provide us with a controlled system to examine the microscopic differences that accompany their distinct transport responses (see Figure~\ref{fig:hydration} and Table~\ref{tab:spd_sites}). We analyze the hydration behavior of the polyamines at $E=\SI{0.135}{\volt\per\nano\meter}$ by looking at the nitrogen atoms' coordination with water molecules. The coordination number is computed as the number of water oxygen atoms within \SI{0.35}{\nano\meter} from the reference atom. The coordination of a particular site is governed almost entirely by its own protonation state and position. Protonated, terminal nitrogens have coordination numbers ranging from 4.0 to 4.4 and neutral ones 2.4 to 2.6, whereas the central nitrogen holds 2.1--2.2 waters when protonated and 1.6--1.8 when neutral. For $1,5\text{-}\mathrm{Spd}^{2+}$, the nitrogen atoms' total coordination number is $N_T=9.12$, compared to $N_T=9.74$ for $1,10\text{-}\mathrm{Spd}^{2+}$, and $N_T=10.43$ for the fully protonated spermidine. The velocity of $1,10\text{-}\mathrm{Spd}^{2+}$ and $\mathrm{Spd}^{3+}$ differs by a factor of 2.2 unveiling a key role of hydration. $\mathrm{Spd}^{3+}$ is the most hydrated and slowest polyamine while carrying the highest charge and displaying a stronger association with chloride ions. Total hydration, however, does not entirely fix the velocity. At fixed charge, $1,10\text{-}\mathrm{Spd}^{2+}$ is slightly more hydrated than $1,5\text{-}\mathrm{Spd}^{2+}$ and has a 42\% higher velocity, with a protonated nitrogen on the terminal site rather than the central one, allowing the charged sites to hold 8.0 rather than 6.5 waters in the case of $1,5\text{-}\mathrm{Spd}^{2+}$. The radii of gyration of the three species differ by about 3\% (3.19, 3.27, and \SI{3.29}{\angstrom}). With comparable radii of gyration, our results suggest that the velocity difference between $1,10\text{-}\mathrm{Spd}^{2+}$ and $1,5\text{-}\mathrm{Spd}^{2+}$ is due to the charge placement. 

Water-density maps shown in Figure~\ref{fig:hydration}a illustrate how each amine locates within the confined water. Away from the solute, water in the $R=\SI{0.68}{\nano\meter}$ CNT is highly structured, appearing as two axial regions of 1.6--1.7 times bulk density. Over an axial window of $\approx\SI{1}{\nano\meter}$ centered on the polyamine, the average water density falls below 80\% of the density far from the solute, and to 30--40\% near the polyamine center of mass. Water accumulates at both ends of the molecule, where local density reaches 1.6--1.9 times the bulk density. Within the excluded region, $\mathrm{Spd}^{3+}$ retains the most water. The polyamines pictured displace nearly the same amount of water in total, differing in how the remaining water distributes among the nitrogen sites (Table~\ref{tab:spd_sites}).

\begin{table}[tbp]
\centering
\caption{Hydration, chloride association, molecular size, and mobility of spermidine protonation states in the $R=\SI{0.68}{\nano\meter}$ CNT at $E=\SI{0.135}{\volt\per\nano\meter}$. According to their position in the molecule, the coordination number of N-atoms with water molecules is designated as $\mathrm{N}^1$, $\mathrm{N}^5$, and $\mathrm{N}^{10}$. Coordination numbers are computed with a cutoff radius of $\SI{0.35}{\nano\meter}$. Protonated sites are marked with an asterisk; they display standard errors $\leq0.04$, and $N_T$ is their sum over the three nitrogens. Cl$^-$ is the mean number of chloride ions within cutoff radius $\SI{0.45}{\nano\meter}$ of any nitrogen; $R_g$ is the radius of gyration (with standard errors uniformly $\leq\SI{0.01}{\angstrom}$). $\mu$ is reported in nm$^2$\,V$^{-1}$\,ns$^{-1}$. Reported values are averages from three independent trajectories.}
\label{tab:spd_sites}
\small
\begin{tabular}{lccccccc}
\toprule
System & $\mathrm{N}^1$ & $\mathrm{N}^5$ & $\mathrm{N}^{10}$ & $N_T$ & Cl$^-$ & $R_g$ (\AA) & $\mu$ \\
\midrule
$\mathrm{Spd}^{0}$ & 2.53  & 1.57  & 2.44  & 6.54  & 0.01 & 3.36 & --- \\
$1\text{-}\mathrm{Spd}^{1+}$ & 4.20* & 1.78  & 2.46  & 8.44  & 0.04 & 3.26 & 6.06 \\
$1{,}5\text{-}\mathrm{Spd}^{2+}$ & 4.39* & 2.13* & 2.60  & 9.12  & 0.59 & 3.19 & 3.05 \\
$1{,}10\text{-}\mathrm{Spd}^{2+}$ & 4.03* & 1.75  & 3.96* & 9.74  & 0.56 & 3.27 & 4.33 \\
$\mathrm{Spd}^{3+}$ & 4.10* & 2.16* & 4.17* & 10.43 & 1.22 & 3.29 & 1.97 \\
\bottomrule
\end{tabular}
\end{table}

Nitrogen--water-oxygen radial density for each site provides an additional local comparison (Figure~\ref{fig:hydration}b). The profiles distinguish hydration between protonated and neutral nitrogen atoms. In each molecular charge distribution, the protonated nitrogens carry a stronger first shell. However, the bracketed differences between the two protonated-nitrogen first peaks demonstrate a discrepancy between the two microspecies---large for $1{,}5\text{-}\mathrm{Spd}^{2+}$, whose two protonated sites are inequivalent, and nearly vanishing for $1{,}10\text{-}\mathrm{Spd}^{2+}$, whose protonated sites are roughly equally hydrated. In $\mathrm{Spd}^{3+}$ the terminal peaks (3.6--3.9) exceed that of the central $\mathrm{N}^5$ (2.3). Relocating a proton while fixing the molecular backbone therefore redistributes hydration between sites. This structural difference accompanies the observed modulation in electrophoretic motion.

Likewise, coordination on terminal-nitrogen sites in the fully protonated amines does not uniquely track the inversion in mobility; at $R=\SI{1.02}{\nano\meter}$, the ordering in mobility is inverted while the multivalent amines are more coordinated than ethylamine (Supporting Information). The neutral CNT wall does not hinder the motion of the amines directly---the mean axial force of the CNT on the solute at $R=\SI{0.68}{\nano\meter}$ is statistically consistent with zero and at most 0.3\% of the electrical force $qeE$ (Supporting Information, Section~S7).

\subsection{Implications for neutral nanofluidic selectivity}
\label{sec:implications}

This observed inversion in charge--velocity ordering distinguishes strong neutral confinement from previously studied routes to charge-selective electrophoresis. Bagchi and Olvera de la Cruz, employing Langevin dynamics of bead-spring polyelectrolytes
between charged planar surfaces, demonstrated that imposed surface charge density was required to achieve separation, whereas little charge discrimination was observed with uncharged, wider walls.\cite{bagchi2020jcp} Our atomistic simulations parallel the latter finding: at $R = \SI{1.97}{\nano\meter}$, $\mu/\mu(q{=}1)$ generally increases with charge, approaching conventional ordering. However, with explicit solvent, we observe at $R = \SI{0.68}{\nano\meter}$, pronounced differential transport that does not require imposed surface charge or functionalization.

Free-solution simulations of the polyamines confirm that the (bulk) ordering of mobility values follows theoretical expectations: mobilities increase monotonically from ethylamine $q=+1$ to spermine $q = +4$. Absolute bulk mobility values ($51$--$\SI{67}{\nano\meter^2/(V \cdot ns)}$) are specific to the model. In free solution, these mobilities rise by 15--26\% between 0.135 and \SI{0.900}{\volt\per\nano\meter}, nearly uniformly for all polyamines (Supporting Information, Section~S6 and Figure~S1f). However, the mobility of $\mathrm{Eta}^{1+}$ increases by a factor of 1.5 in the narrowest CNT while that of $\mathrm{Spm}^{4+}$ rises fourfold over the same range of electric field strengths (Figure~\ref{fig:resistance}a and \ref{fig:nonlinear}a). This frames the strong field dependence of mobility observed in the CNTs as a consequence specific to confinement.

In angstrom-scale pores, ion selectivity can be associated with partial dehydration and with free-energy costs of disrupting hydration shells upon pore entry.\cite{tunuguntla2017} The polyamines simulated remain hydrated inside the CNT, so the present response is not a simple dehydration barrier. Rather, confinement imposes geometric restrictions on the coordination environments of protonation sites. The comparison between $q=+2$ spermidine microspecies illustrates how the relocation of charge on a molecular scaffold influences both structure of hydration and transport overall.

Further, spermidine macro- and microspecies simulations show that net charge is not the sole molecular variable which emerges in the transport response. This indicates that strongly confined neutral transport can resolve both charge state and the spatial distribution of charge.

Because polyamine protonation states depend on pH, the spermidine macrospecies data suggest a possible route to tunable nanofluidic transport; however, protonation equilibria under confinement were not calculated, so the present results do not predict a specific-pH dependent separation. Likewise, $q=+2$ spermidine microspecies demonstrate a sensitivity to charge topology, which could correspond to a readable kinetic coordinate, but do not alone demonstrate a direct residue-by-residue sequence readout. 

These simulations employ fixed solute and ion charges, SPC/E water, and rigid, nonpolarizable CNTs. Thus, quantitative transport rates are model specific, and polarization, charge transfer, and CNT phonon-mediated energy dissipation are absent. Wall polarization, which modifies correlations and conductivity of ions in confinement,\cite{dossantos2023polarizable,jimenez2023polarization} would also alter association of counterions observed. Because the CNT is axially periodic, these simulations give differential mobility within the CNT; entry kinetics, barrier effects, and finite-channel-length effects are not modeled here. These approximations limit direct application of absolute mobilities and force partitioning to experimental CNTs, leaving internally controlled comparisons, such as spermidine microspecies, well defined within this simulation model.

\section{Conclusions}
\label{sec:conclusions}

All-atom non-equilibrium molecular dynamics simulations reveal that uncharged, geometrically confining carbon nanotubes can reverse the conventional valence-dependent electrophoretic mobility. When subjected to strong confinement, spermine migrates more slowly than ethylamine, despite its greater charge. The mobility falls faster than $1/q$ for the parent amines, so the altered transport response is not captured by charge alone. This inversion diminishes as confinement is relaxed, and the mobility ordering in free solution is recovered. Neutral geometric confinement can therefore generate pronounced valence-selective transport without imposed surface charge. Transport properties among confined spermidine macrospecies show that, at low field, increasing protonation corresponds to a strong decrease in mobility. Fixed-charge microspecies further show that the response is sensitive to not only net valence, but protonation topology. Moreover, the two dicationic spermidine microspecies show distinct transport responses and hydration structures across individual nitrogens. The hydration analysis identifies charge placement as a contributing factor to local solvation behavior.

\section{Methods}
\label{sec:methods}

\subsection{System construction}
\label{sec:methods:system}

We generated the CNTs using the Atomic Simulation Environment with a carbon--carbon bond length of $\SI{1.42}{\mathring{A}}$ and with axes in alignment with $z$. We assigned CNT carbon atoms the AMBER03 sp$^2$ carbon Lennard-Jones parameters,\cite{duan2003} $\sigma = \SI{3.40e-1}{\nano\meter}$, $\epsilon = \SI{3.60e-1}{kJ/mol}$. Cross-interactions were set by Lorentz--Berthelot combination rules. CNT carbon atom positions were frozen in all three dimensions. Equilibration coupled to a reservoir provides a reference for background water density; this procedure is outlined in the SI.

Periodic boundary conditions were applied in $x$, $y$, and $z$ dimensions within an orthorhombic unit cell with axial length and discrete KCl count chosen to obtain approximately \SI{50}{mM} added electrolyte. Each system was solvated with SPC/E water molecules\cite{berendsen1987} using the GROMACS \texttt{gmx solvate} utility, with water retained only within the CNT lumen and the exterior as vacuum. Complete cell dimensions and system compositions are reported in the SI.

\subsection{Polyamines}
\label{sec:methods:polyamines}

We simulate four polyamines---ethylamine (Eta), putrescine (Put),
spermidine (Spd), and spermine (Spm)---in their fully protonated states, with net charges of $+1e$, $+2e$, $+3e$, and $+4e$, respectively. Intramolecular parameters including bonds, angles, and dihedrals, as well as atomic partial
charges, use ACPYPE, General AMBER Force Field
(GAFF) for bonding and Lennard--Jones parameters, as well as
AM1-BCC for partial charges.\cite{acpype,wang2004,jakalian2002} Additionally, cross-interactions between polyamine atoms and ion/water atoms follow  Lorentz--Berthelot combination rules.

To characterize the competing contributions from valence and molecular size, we constructed a series of macrospecies, differing in protonation state, of spermidine, denoted as
$\text{Spd}^0$ ($q = 0$),
$\text{1-Spd}^{1+}$ ($q = +1$),
$\text{1,5-Spd}^{2+}$ ($q = +2$, protonation at N$^1$ and N$^5$),
$\text{1,10-Spd}^{2+}$ ($q = +2$, protonation at N$^1$ and N$^{10}$), and
$\text{Spd}^{3+}$ ($q = +3$, fully protonated),
each with the same polyamine backbone. All polyamines and spermidine macrospecies reported employ GAFF/AM1-BCC parameters, with protonated nitrogens assigned the GAFF \texttt{n4} type and neutral nitrogens the \texttt{n3} type. The Supporting Information identifies the topology and trajectories used for each parent amine and spermidine protonation variant.

\subsection{Force fields and simulation parameters}
\label{sec:methods:ff}

We performed all simulations using GROMACS 2024.5,\cite{abraham2015} with SPC/E as the water
model,\cite{berendsen1987}
with Lennard-Jones parameters for
K$^+$,\cite{aqvist1990} and Cl$^-$ parameters from Smith and
Dang\cite{smithdang1994} included in the AMBER03
FF\cite{duan2003} and combined through Lorentz--Berthelot rules.

We treated long-range electrostatics using the particle-mesh Ewald (PME)
method\cite{essmann1995} with real-space cutoff $\SI{1.0}{\nano\meter}$, Fourier grid spacing $\SI{0.12}{\nano\meter}$, and fourth-order interpolation. We truncate van der Waals interactions at $\SI{1.0}{\nano\meter}$ using the Verlet cutoff scheme with neighbor-list update frequency of 20 timesteps.

The velocity-rescale thermostat maintained the temperature at $\SI{300}{K}$ \cite{bussi2007} with a coupling time constant of $\SI{1.0}{ps}$ on the entire system as a singular thermostat group. Simulations were performed at fixed volume due to the rigid CNT geometry, and without pressure coupling. Linear center-of-mass motion was removed every 100 steps as per GROMACS default \texttt{comm-mode = Linear} and \texttt{nstcomm = 100}. The effects of the thermostat, periodic images, and cell length were assessed (Supporting Information). We constrained covalent bonds with the LINCS algorithm\cite{hess1997}
(\texttt{lincs-order $= 4$}, \texttt{lincs-iter $= 1$)}; the simulation timestep was $\SI{2}{fs}$ for both equilibration and production, without hydrogen-mass repartitioning.

\subsection{Simulation Production}
\label{sec:methods:nemd}

We equilibrate each system for $\SI{10}{ns}$ under NVT conditions at $\SI{300}{K}$ with no applied electric field. This equilibration uses a \SI{2}{fs} timestep and thermostat coupling time \SI{0.5}{ps}. GROMACS \texttt{electric-field-z} imposes a uniform field which applies a force $\mathbf{F}_i=q_i\mathbf{E}$ directly to each particle. Including the zero-field run, field strengths range from $0.00$ to $\SI{0.150}{\volt\per\nano\meter}$ in $\SI{0.015}{\volt\per\nano\meter}$ increments for all CNT radii.
We perform additional simulations of electric field strengths up to $\SI{0.900}{\volt\per\nano\meter}$, each for $\SI{500}{ns}$ with coordinates saved at $\SI{20}{ps}$ intervals, to test the persistence of the observed velocity ordering beyond low electric field strengths. 

For all fully protonated amines at all radii and for the spermidine charge variants at radius $R = \SI{0.68}{\nano\meter}$ (all at $E = \SI{0.135}{V\per \nano\meter}$), we perform two additional replicas, per amine, initiated from one equilibrated system configuration, in order to estimate statistical uncertainty. Three $\SI{500}{\nano\second}$ trajectories thus comprise the mean for parent-amine and spermidine-variant data reported at this field.

\subsection{Free-solution simulations}
\label{sec:methods:bulk}

We establish a baseline reference for polyamine transport in the absence of confinement by simulating the polyamines in free solution with the same force-field and thermostat conditions. In each system, we place one polyamine in a $\SI{5}{\nano\meter}$ cubic box of SPC/E water with approximately $\SI{50}{mM}$ KCl. We sample $E=0$ and $\SI{0.135}{\volt\per\nano\meter}$ with three $\SI{500}{ns}$ trajectories each, and 0.300, 0.600, and \SI{0.900}{\volt\per\nano\meter} with one trajectory each. 

\subsection{Analysis}
\label{sec:methods:analysis}

We perform all simulation analysis using MDAnalysis.\cite{michaudagrawal2011,gowers2016} Drift velocities were computed using GROMACS \texttt{trjconv -pbc nojump} to obtain continuous coordinates across the periodic boundaries. For each trajectory, the axial coordinate was obtained as the polyamine center of mass for amines, the mean axial coordinate for water oxygens, and the mean axial coordinate of each ion species for $\text{K}^+$ and $\text{Cl}^-$. 
With each trajectory, we estimated the mean axial drift velocity by fitting the unwrapped axial coordinate via least-squares regression,
\begin{equation}
  z_\alpha(t) = z_{\alpha,0}+v_\alpha t.
  \label{eq:drift}
\end{equation}
The first \SI{5}{\nano\second} of each trajectory was discarded; the Supporting Information reports the individual velocities of each replicate.

Electrophoretic mobility of species $\alpha$ was computed as
\begin{equation}
  \mu = \frac{v_{\mathrm{amine}}}{E}\,.
  \label{eq:mobility}
\end{equation}
For polyamines, mobilities were normalized by the mobility of ethylamine measured at the
same radius and field,
\begin{equation}
  \frac{\mu(q,R,E)}{\mu(q{=}1,R,E)}\,,
  \label{eq:mu_norm}
\end{equation}
to compare the relative velocities of the differently charged amines.

Water coordination was computed as the number of water oxygen atoms within a cutoff radius $r_c$ of each nitrogen.
Coordination numbers were evaluated over protonated and neutral nitrogens for each spermidine charge variant.

Nitrogen and water-oxygen radial densities were compared by counting water oxygens in spherical shells centered on each nitrogen and normalizing by the shell volume, the number of sampled frames, and by the SPC/E bulk water-oxygen density (\SI{33.4}{\per\nano\meter\cubed}). Given that the confined liquid is spatially inhomogeneous and shells may intersect the CNT wall, these data are not interpreted as explicit radial distribution functions. 

Two-dimensional water density maps were computed in an amine-centered frame from the three trajectories  at $E=\SI{0.135}{\volt\per\nano\meter}$ for the
spermidine charge variants in the CNT of radius $\SI{0.68}{\nano\meter}$.
At each sampled frame, water oxygen coordinates were translated by the amine center of mass.
Computed from a $y$-layer of thickness $\SI{3.0}{\angstrom}$, densities were plotted on an
$x$--$z$ grid with $\SI{0.3}{\angstrom}$ spacing, normalized over the sampled volume and SPC/E bulk water density.

\subsection{Hydration analysis}
\label{sec:methods:hydration}

We analyzed three \SI{500}{\nano\second} trajectories of two dicationic spermidine microspecies and of $\mathrm{Spd}^{3+}$ at $E=\SI{0.135}{\volt\per\nano\meter}$, each sampled at \SI{0.1}{\nano\second} intervals, discarding the first \SI{50}{\nano\second}. The Supporting Information further specifies the system and simulation protocols. We characterized hydration at each nitrogen site by nitrogen--water-oxygen distance distributions and first-shell coordination numbers. For nitrogen site $i$, the instantaneous coordination number is
\begin{equation}
    N_i(t) = \sum_{j\in\mathrm{water}}
    \mathbf{1}\!\left[r_{ij}(t) < r_c\right],
    \qquad r_c = \SI{0.35}{\nano\meter},
    \label{eq:cn}
\end{equation}
where $r_{ij}$ is the minimum-image distance between nitrogen $i$ and water oxygen $j$; $\langle N_i\rangle$ was evaluated for each site across the replicate trajectories before averaging, and the reported uncertainties are standard errors of the means. Site labels follow the nitrogen-position notation $\mathrm{N}^1$, $\mathrm{N}^5$, and $\mathrm{N}^{10}$ of spermidine. 

Association with counterions was quantified as a supporting measure, calculated as the number of chloride ions within \SI{0.45}{\nano\meter} from a nitrogen atom relative to the uniform distribution (Supporting Information). To quantify the field dependence (Figure~\ref{fig:nonlinear}b,c), chloride coordination and $R_g$ were sampled every 0.5 and \SI{1}{\nano\second}, respectively. The first \SI{50}{\nano\second} of each simulation trajectory was discarded.

\section*{Supporting Information}
The Supporting Information contains the complete system-build protocols; transport data for each simulation trial; controls for system construction, solvation, and transport methodology; counterion association values and CNT-wall force; and hydration-structure metrics as well as the dependence of hydration data on the selected cutoff value. 

\begin{acknowledgement}
This work was supported by the National Science Foundation through the Designing Materials to Revolutionize and Engineer our Future (DMREF) program (Grant No.~DMR-2323316) and the Northwestern Materials Research Science and Engineering Center (MRSEC) Research Experience for Undergraduates (REU) program (Grant No.~DMR-2308691). We also thank QUEST at Northwestern University for providing computational facilities.
\end{acknowledgement}

\section*{Conflict of Interest}
The authors declare no competing financial interest.

\section*{Data Availability}
The data that support the findings of this study are available
from the corresponding author upon reasonable request.

\bibliography{references}

@article{kasianowicz1996,
    author  = {Kasianowicz, John J. and Brandin, Eric and Branton, Daniel and Deamer, David W.},
    title   = {Characterization of individual polynucleotide molecules using a membrane channel},
    journal = {Proc. Natl. Acad. Sci. U.S.A.},
    volume  = {93},
    number  = {24},
    pages   = {13770--13773},
    year    = {1996},
    doi     = {10.1073/pnas.93.24.13770},
}

@article{manrao2012,
    author  = {Manrao, Elizabeth A. and Derrington, Ian M. and Laszlo, Andrew H. and Langford, Kyle W. and Hopper, Matthew K. and Gillgren, Nathaniel and Pavlenok, Mikhail and Niederweis, Michael and Gundlach, Jens H.},
    title   = {Reading {DNA} at single-nucleotide resolution with a mutant {MspA} nanopore and phi29 {DNA} polymerase},
    journal = {Nat. Biotechnol.},
    volume  = {30},
    number  = {4},
    pages   = {349--353},
    year    = {2012},
    doi     = {10.1038/nbt.2171},
}

@article{tunuguntla2017,
    author  = {Tunuguntla, Ramya H. and Henley, Robert Y. and Yao, Yun-Chiao and Pham, Tuan Anh and Wanunu, Meni and Noy, Aleksandr},
    title   = {Enhanced water permeability and tunable ion selectivity in subnanometer carbon nanotube porins},
    journal = {Science},
    volume  = {357},
    number  = {6353},
    pages   = {792--796},
    year    = {2017},
    doi     = {10.1126/science.aan2438},
}

@book{hunter1981,
    author    = {Hunter, Robert J.},
    title     = {Zeta Potential in Colloid Science: Principles and Applications},
    publisher = {Academic Press},
    address   = {London},
    year      = {1981},
}

@article{bocquet2010,
    author  = {Bocquet, Lyd{\'e}ric and Charlaix, Elisabeth},
    title   = {Nanofluidics, from bulk to interfaces},
    journal = {Chem. Soc. Rev.},
    volume  = {39},
    number  = {3},
    pages   = {1073--1095},
    year    = {2010},
    doi     = {10.1039/B909366B},
}

@article{sparreboom2009,
    author  = {Sparreboom, Wouter and van den Berg, Albert and Eijkel, Jan C. T.},
    title   = {Principles and applications of nanofluidic transport},
    journal = {Nat. Nanotechnol.},
    volume  = {4},
    number  = {11},
    pages   = {713--720},
    year    = {2009},
    doi     = {10.1038/nnano.2009.332},
}

@article{hummer2001,
    author  = {Hummer, Gerhard and Rasaiah, Jayendran C. and Noworyta, Jerzy P.},
    title   = {Water conduction through the hydrophobic channel of a carbon nanotube},
    journal = {Nature},
    volume  = {414},
    number  = {6860},
    pages   = {188--190},
    year    = {2001},
    doi     = {10.1038/35102535},
}

@article{holt2006,
    author  = {Holt, Jason K. and Park, Hyung Gyu and Wang, Yinmin and Stadermann, Michael and Artyukhin, Alexander B. and Grigoropoulos, Costas P. and Noy, Aleksandr and Bakajin, Olgica},
    title   = {Fast Mass Transport Through Sub-2-Nanometer Carbon Nanotubes},
    journal = {Science},
    volume  = {312},
    number  = {5776},
    pages   = {1034--1037},
    year    = {2006},
    doi     = {10.1126/science.1126298},
}

@article{majumder2005,
    author  = {Majumder, Mainak and Chopra, Nitin and Andrews, Rodney and Hinds, Bruce J.},
    title   = {Enhanced flow in carbon nanotubes},
    journal = {Nature},
    volume  = {438},
    number  = {7064},
    pages   = {44},
    year    = {2005},
    doi     = {10.1038/438044a},
}

@article{secchi2016,
    author  = {Secchi, Eleonora and Marbach, Sophie and Niguès, Antoine and Stein, Derek and Siria, Alessandro and Bocquet, Lydéric},
    title   = {Massive radius-dependent flow slippage in carbon nanotubes},
    journal = {Nature},
    volume  = {537},
    number  = {7619},
    pages   = {210--213},
    year    = {2016},
    doi     = {10.1038/nature19315},
}

@article{bagchi2020jcp,
    author  = {Bagchi, Debarshee and Olvera de la Cruz, Monica},
    title   = {Dynamics of a driven confined polyelectrolyte solution},
    journal = {J. Chem. Phys.},
    volume  = {153},
    number  = {18},
    pages   = {184904},
    year    = {2020},
    doi     = {10.1063/5.0027049},
}

@article{hoagland1999,
    author  = {Hoagland, David A. and Arvanitidou, Evangelia and Welch, Cynthia},
    title   = {Capillary electrophoresis measurements of the free solution mobility for several model polyelectrolyte systems},
    journal = {Macromolecules},
    volume  = {32},
    number  = {19},
    pages   = {6180--6190},
    year    = {1999},
    doi     = {10.1021/ma9903761},
}

@article{stellwagen1997,
    author  = {Stellwagen, Nancy C. and Gelfi, Cecilia and Righetti, Pier Giorgio},
    title   = {The free solution mobility of {DNA}},
    journal = {Biopolymers},
    volume  = {42},
    number  = {6},
    pages   = {687--703},
    year    = {1997},
    doi     = {10.1002/(SICI)1097-0282(199711)42:6<687::AID-BIP7>3.0.CO;2-Q},
}

@article{tabor1984,
    author  = {Tabor, Celia White and Tabor, Herbert},
    title   = {Polyamines},
    journal = {Annu. Rev. Biochem.},
    volume  = {53},
    pages   = {749--790},
    year    = {1984},
    doi     = {10.1146/annurev.bi.53.070184.003533},
}

@article{igarashi2010,
    author  = {Igarashi, Kazuei and Kashiwagi, Keiko},
    title   = {Modulation of cellular function by polyamines},
    journal = {Int. J. Biochem. Cell Biol.},
    volume  = {42},
    number  = {1},
    pages   = {39--51},
    year    = {2010},
    doi     = {10.1016/j.biocel.2009.07.009},
}

@article{pegg2016,
    author  = {Pegg, Anthony E.},
    title   = {Functions of Polyamines in Mammals},
    journal = {J. Biol. Chem.},
    volume  = {291},
    number  = {29},
    pages   = {14904--14912},
    year    = {2016},
    doi     = {10.1074/jbc.R116.731661},
}

@article{casero2018,
    author  = {Casero, Robert A. and Stewart, Tracy Murray and Pegg, Anthony E.},
    title   = {Polyamine metabolism and cancer: treatments, challenges and opportunities},
    journal = {Nat. Rev. Cancer},
    volume  = {18},
    number  = {11},
    pages   = {681--695},
    year    = {2018},
    doi     = {10.1038/s41568-018-0050-3},
}

@article{abraham2015,
    author  = {Abraham, Mark James and Murtola, Teemu and Schulz, Roland and P{\'a}ll, Szil{\'a}rd and Smith, Jeremy C. and Hess, Berk and Lindahl, Erik},
    title   = {{GROMACS}: High performance molecular simulations through multi-level parallelism from laptops to supercomputers},
    journal = {SoftwareX},
    volume  = {1--2},
    pages   = {19--25},
    year    = {2015},
    doi     = {10.1016/j.softx.2015.06.001},
}

@article{berendsen1987,
    author  = {Berendsen, H. J. C. and Grigera, J. R. and Straatsma, T. P.},
    title   = {The missing term in effective pair potentials},
    journal = {J. Phys. Chem.},
    volume  = {91},
    number  = {24},
    pages   = {6269--6271},
    year    = {1987},
    doi     = {10.1021/j100308a038},
}

@article{acpype,
    author  = {Sousa da Silva, Alan W. and Vranken, Wim F.},
    title   = {{ACPYPE} --- {A}nte{C}hamber {PY}thon {P}arser interfac{E}},
    journal = {BMC Res. Notes},
    volume  = {5},
    pages   = {367},
    year    = {2012},
    doi     = {10.1186/1756-0500-5-367},
}

@article{wang2004,
    author  = {Wang, Junmei and Wolf, Romain M. and Caldwell, James W. and Kollman, Peter A. and Case, David A.},
    title   = {Development and testing of a general amber force field},
    journal = {J. Comput. Chem.},
    volume  = {25},
    number  = {9},
    pages   = {1157--1174},
    year    = {2004},
    doi     = {10.1002/jcc.20035},
}

@article{jakalian2002,
    author  = {Jakalian, Araz and Jack, David B. and Bayly, Christopher I.},
    title   = {Fast, efficient generation of high-quality atomic charges. {AM1-BCC} model: {II}. {P}arameterization and validation},
    journal = {J. Comput. Chem.},
    volume  = {23},
    number  = {16},
    pages   = {1623--1641},
    year    = {2002},
    doi     = {10.1002/jcc.10128},
}

@article{duan2003,
    author  = {Duan, Yong and Wu, Chun and Chowdhury, Shibasish and Lee, Mathew C. and Xiong, Guoming and Zhang, Wei and Yang, Rong and Cieplak, Piotr and Luo, Ray and Lee, Taisung and Caldwell, James and Wang, Junmei and Kollman, Peter},
    title   = {A point-charge force field for molecular mechanics simulations of proteins based on condensed-phase quantum mechanical calculations},
    journal = {J. Comput. Chem.},
    volume  = {24},
    number  = {16},
    pages   = {1999--2012},
    year    = {2003},
    doi     = {10.1002/jcc.10349},
}

@article{essmann1995,
    author  = {Essmann, Ulrich and Perera, Lalith and Berkowitz, Max L. and Darden, Tom and Lee, Hsing and Pedersen, Lee G.},
    title   = {A smooth particle mesh {E}wald method},
    journal = {J. Chem. Phys.},
    volume  = {103},
    number  = {19},
    pages   = {8577--8593},
    year    = {1995},
    doi     = {10.1063/1.470117},
}

@article{aqvist1990,
    author  = {{\AA}qvist, Johan},
    title   = {Ion-water interaction potentials derived from free energy perturbation simulations},
    journal = {J. Phys. Chem.},
    volume  = {94},
    number  = {21},
    pages   = {8021--8024},
    year    = {1990},
    doi     = {10.1021/j100384a009},
}

@article{smithdang1994,
    author  = {Smith, David E. and Dang, Liem X.},
    title   = {Computer simulations of {NaCl} association in polarizable water},
    journal = {J. Chem. Phys.},
    volume  = {100},
    number  = {5},
    pages   = {3757--3766},
    year    = {1994},
    doi     = {10.1063/1.466363},
}

@article{bussi2007,
    author  = {Bussi, Giovanni and Donadio, Davide and Parrinello, Michele},
    title   = {Canonical sampling through velocity rescaling},
    journal = {J. Chem. Phys.},
    volume  = {126},
    number  = {1},
    pages   = {014101},
    year    = {2007},
    doi     = {10.1063/1.2408420},
}

@article{hess1997,
    author  = {Hess, Berk and Bekker, Henk and Berendsen, Herman J. C. and Fraaije, Johannes G. E. M.},
    title   = {{LINCS}: A linear constraint solver for molecular simulations},
    journal = {J. Comput. Chem.},
    volume  = {18},
    number  = {12},
    pages   = {1463--1472},
    year    = {1997},
    doi     = {10.1002/(SICI)1096-987X(199709)18:12<1463::AID-JCC4>3.0.CO;2-H},
}

@article{michaudagrawal2011,
    author  = {Michaud-Agrawal, Naveen and Denning, Elizabeth J. and Woolf, Thomas B. and Beckstein, Oliver},
    title   = {{MDA}nalysis: A toolkit for the analysis of molecular dynamics simulations},
    journal = {J. Comput. Chem.},
    volume  = {32},
    number  = {10},
    pages   = {2319--2327},
    year    = {2011},
    doi     = {10.1002/jcc.21787},
}

@inproceedings{gowers2016,
    author    = {Gowers, Richard J. and Linke, Max and Barnoud, Jonathan and Reddy, Tyler J. E. and Melo, Manuel N. and Seyler, Sean L. and Domanski, Jan and Dotson, David L. and Buchoux, S{\'e}bastien and Kenney, Ian M. and Beckstein, Oliver},
    title     = {{MDA}nalysis: A {P}ython package for the rapid analysis of molecular dynamics simulations},
    booktitle = {Proceedings of the 15th Python in Science Conference},
    editor    = {Benthall, Sebastian and Rostrup, Scott},
    pages     = {98--105},
    year      = {2016},
    doi       = {10.25080/Majora-629e541a-00e},
}

@article{luan2008,
    author  = {Luan, Binquan and Aksimentiev, Aleksei},
    title   = {Electro-osmotic screening of the {DNA} charge in a nanopore},
    journal = {Phys. Rev. E},
    volume  = {78},
    number  = {2},
    pages   = {021912},
    year    = {2008},
    doi     = {10.1103/PhysRevE.78.021912},
}

@article{shendruk2015,
    author  = {Shendruk, Tyler N. and Bertrand, Martin and Slater, Gary W.},
    title   = {Electrophoretic Mobility of Polyelectrolytes within a Confining Well},
    journal = {ACS Macro Lett.},
    volume  = {4},
    number  = {4},
    pages   = {472--476},
    year    = {2015},
    doi     = {10.1021/acsmacrolett.5b00076},
}

@article{fong2025,
    author  = {Fong, Kara D. and Grey, Clare P. and Michaelides, Angelos},
    title   = {On the Physical Origins of Reduced Ionic Conductivity in Nanoconfined Electrolytes},
    journal = {ACS Nano},
    volume  = {19},
    number  = {13},
    pages   = {13191--13201},
    year    = {2025},
    doi     = {10.1021/acsnano.4c18956},
}

@article{zhang2026hydration,
    author  = {Zhang, Yuyang and Zhang, Ziyin and Peng, Haoke and Li, Qizhu and Zou, Yi-Chao and Wang, Deqiang and Wu, Heng-An and Wang, Fengchao},
    title   = {Bridging Hydration-Shell Exchange Kinetics and Ion Translocation Energy Barriers across Graphene Nanopores},
    journal = {Nano Lett.},
    volume  = {26},
    pages   = {3188--3194},
    year    = {2026},
    doi     = {10.1021/acs.nanolett.5c06456},
}

@article{dossantos2023polarizable,
  author  = {dos Santos, Alexandre P. and Jim{\'e}nez-{\'A}ngeles, Felipe and Ehlen, Ali and Olvera de la Cruz, Monica},
  title   = {Modulation of ionic conduction using polarizable surfaces},
  journal = {Phys. Rev. Res.},
  volume  = {5},
  pages   = {043174},
  year    = {2023},
  doi     = {10.1103/PhysRevResearch.5.043174}
}

@article{jimenez2023polarization,
  author  = {Jim{\'e}nez-{\'A}ngeles, Felipe and Ehlen, Ali and Olvera de la Cruz, Monica},
  title   = {Surface polarization enhances ionic transport and correlations in electrolyte solutions nanoconfined by conductors},
  journal = {Faraday Discuss.},
  volume  = {246},
  pages   = {576--591},
  year    = {2023},
  doi     = {10.1039/D3FD00028A}
}

@article{motone2024multipass,
  author  = {Motone, Keisuke and Kontogiorgos-Heintz, Daphne and Wee, Jasmine and Kurihara, Kyoko and Yang, Sangbeom and Roote, Gwendolin and Fox, Oren E. and Fang, Yishu and Queen, Melissa and Tolhurst, Mattias and Cardozo, Nicolas and Jain, Miten and Nivala, Jeff},
  title   = {Multi-pass, single-molecule nanopore reading of long protein strands},
  journal = {Nature},
  volume  = {633},
  pages   = {662--669},
  year    = {2024},
  doi     = {10.1038/s41586-024-07935-7}
}

@article{wang2023amino,
  author  = {Wang, Fushi and Zhao, Chunxiao and Zhao, Pinlong and Chen, Fanfan and Qiao, Dan and Feng, Jiandong},
  title   = {{MoS$_2$} nanopore identifies single amino acids with sub-1 {Dalton} resolution},
  journal = {Nat. Commun.},
  volume  = {14},
  pages   = {2895},
  year    = {2023},
  doi     = {10.1038/s41467-023-38627-x}
}

@article{shen2025cnt,
  author  = {Shen, Yingjun and Ding, Mingming},
  title   = {Discrimination and Translocation of Charged Proteinogenic Amino Acids through a Single-Walled Carbon Nanotube},
  journal = {J. Phys. Chem. B},
  volume  = {129},
  number  = {13},
  pages   = {3502--3513},
  year    = {2025},
  doi     = {10.1021/acs.jpcb.4c08692}
}

@article{sur2025orientation,
  author  = {Sur, Pranjal and Upadhyaya, Anurag and Varma, Manoj and Maiti, Prabal K.},
  title   = {Orientation dependence of current blockade in single amino acid translocation through a graphene nanopore},
  journal = {Nanoscale},
  volume  = {17},
  pages   = {11016--11027},
  year    = {2025},
  doi     = {10.1039/D4NR04630G}
}

@article{bonthuis2012permittivity,
  author  = {Bonthuis, Douwe Jan and Gekle, Stephan and Netz, Roland R.},
  title   = {Profile of the Static Permittivity Tensor of Water at Interfaces: Consequences for Capacitance, Hydration Interaction and Ion Adsorption},
  journal = {Langmuir},
  volume  = {28},
  pages   = {7679--7694},
  year    = {2012},
  doi     = {10.1021/la2051564}
}

@article{olvera1986electrophoresis,
  author  = {Olvera de la Cruz, M. and Deutsch, J. M. and Edwards, S. F.},
  title   = {Electrophoresis in strong fields},
  journal = {Phys. Rev. A},
  volume  = {33},
  pages   = {2047},
  year    = {1986},
  doi     = {10.1103/PhysRevA.33.2047}
}

@article{olvera1990pulsed,
  author  = {Olvera de la Cruz, M. and Gersappe, D. and Shaffer, E. O.},
  title   = {Dynamics of {DNA} during Pulsed Field Gel Electrophoresis},
  journal = {Phys. Rev. Lett.},
  volume  = {64},
  pages   = {2324},
  year    = {1990},
  doi     = {10.1103/PhysRevLett.64.2324}
}

@article{robin2021,
  author  = {Robin, Paul and Kavokine, Nikita and Bocquet, Lyd{\'e}ric},
  title   = {Modeling of emergent memory and voltage spiking in ionic transport through angstrom-scale slits},
  journal = {Science},
  volume  = {373},
  pages   = {687--691},
  year    = {2021},
  doi     = {10.1126/science.abf7923}
}

@article{onsager1934,
  author  = {Onsager, Lars},
  title   = {Deviations from {Ohm}'s Law in Weak Electrolytes},
  journal = {J. Chem. Phys.},
  volume  = {2},
  pages   = {599--615},
  year    = {1934},
  doi     = {10.1063/1.1749541}
}

@article{gonzalezmozuelos1995,
  author  = {Gonz{\'a}lez-Mozuelos, Pedro and Olvera de la Cruz, Monica},
  title   = {Ion condensation in salt-free dilute polyelectrolyte solutions},
  journal = {J. Chem. Phys.},
  volume  = {103},
  pages   = {3145--3157},
  year    = {1995}
}

@article{stevens1995,
  author  = {Stevens, Mark J. and Kremer, Kurt},
  title   = {The nature of flexible linear polyelectrolytes in salt free solution: A molecular dynamics study},
  journal = {J. Chem. Phys.},
  volume  = {103},
  pages   = {1669--1690},
  year    = {1995}
}

@article{li2016,
  author  = {Li, Honghao and Erba{\c{s}}, Aykut and Zwanikken, Jos and Olvera de la Cruz, Monica},
  title   = {Ionic Conductivity in Polyelectrolyte Hydrogels},
  journal = {Macromolecules},
  volume  = {49},
  pages   = {9239--9246},
  year    = {2016},
  doi     = {10.1021/acs.macromol.6b01276}
}

@article{raspaud1998,
  author  = {Raspaud, E. and Olvera de la Cruz, M. and Sikorav, J.-L. and Livolant, F.},
  title   = {Precipitation of {DNA} by polyamines: a polyelectrolyte behavior},
  journal = {Biophys. J.},
  volume  = {74},
  number  = {1},
  pages   = {381--393},
  year    = {1998},
  doi     = {10.1016/S0006-3495(98)77795-1}
}

\end{document}


\maketitle

\section{System compositions and trajectories}
\label{si:inventory}

All simulated systems reported in the main text employ a single GAFF/AM1-BCC parametrization; protonated nitrogens are assigned the GAFF \texttt{n4} type, and neutral nitrogens are assigned the \texttt{n3} type. Added KCl in each system met approximately \SI{50}{mM} by choosing the axial cell length such that an integer ion count can result in the desired concentration. The number of neutralizing chloride counterions varies with solute charge.

The salt concentration is defined as
\begin{equation}
    c_{\mathrm{KCl}} = \frac{N_{\mathrm{K}}}{N_{\mathrm{A}} V_{\mathrm{acc}}},
    \qquad
    V_{\mathrm{acc}} = \pi \left(R - \SI{0.17}{\nano\meter} \right)^2 L_z.
    \label{eq:salt}
\end{equation}
Here, $R$ is the geometric CNT radius measured from the atomic coordinates, and \SI{0.17}{\nano\meter} is the van der Waals radius of carbon. The density of water confined in the CNT is $\rho= N_{\mathrm{OW}}/V_{\mathrm{acc}}$, compared to the SPC/E\cite{berendsen1987} bulk value of \SI{33.4}{\per\nano\meter\cubed} at \SI{300}{K}.

\begin{table*}[tbp]
\centering
\caption{System compositions. $L_x$, $L_y$, and $L_z$ are the cell dimensions; $N_{\mathrm{wat}}$ is the number of water molecules inside the CNT, equivalent to the number of water molecules in the system; $c_{\mathrm{KCl}}$ follows Eq.~\ref{eq:salt}; $\rho$ is the density of water in confinement; $d_{\mathrm{CC}}^{\min}$ represents the minimum minimum-image distance between two carbons in the CNT. Axial length decreases with radius because the salt concentration is held near \SI{50}{mM} with integer ion counts.}
\label{tab:Sinventory}
\small
\begin{tabular}{l l c c c c c c c c c}
    \toprule
    $R$ (nm) & System & $L_x$ & $L_y$ & $L_z$ (nm)
      & $N_{\mathrm{wat}}$ & $N_{\mathrm{K}}$ & $N_{\mathrm{Cl}}$
      & $c_{\mathrm{KCl}}$ (mM) & $\rho$ (nm$^{-3}$)
      & $d_{\mathrm{CC}}^{\min}$ (nm) \\
    \midrule
    0.68 & $\mathrm{Eta}^{1+}$ & 2.720 & 2.356 & 39.352 &  944 & 1 & 2 & 52.0 & 29.59 & 0.1411 \\
    0.68 & $\mathrm{Put}^{2+}$ & 2.720 & 2.356 & 39.352 &  960 & 1 & 3 & 52.0 & 30.09 & 0.1411 \\
    0.68 & $\mathrm{Spd}^{3+}$ & 2.720 & 2.356 & 39.352 &  956 & 1 & 4 & 52.0 & 29.96 & 0.1411 \\
    0.68 & $\mathrm{Spm}^{4+}$ & 2.720 & 2.356 & 39.352 &  892 & 1 & 5 & 52.0 & 27.96 & 0.1411 \\
    \midrule
    1.02 & $\mathrm{Eta}^{1+}$ & 3.503 & 3.034 & 29.514 & 2038 & 2 & 3 & 49.9 & 30.64 & 0.1407 \\
    1.02 & $\mathrm{Put}^{2+}$ & 3.503 & 3.034 & 29.514 & 2033 & 2 & 4 & 49.9 & 30.56 & 0.1407 \\
    1.02 & $\mathrm{Spd}^{3+}$ & 3.503 & 3.034 & 29.514 & 2032 & 2 & 5 & 49.9 & 30.55 & 0.1407 \\
    1.02 & $\mathrm{Spm}^{4+}$ & 3.503 & 3.034 & 29.514 & 2023 & 2 & 6 & 49.9 & 30.41 & 0.1407 \\
    \midrule
    1.49 & $\mathrm{Eta}^{1+}$ & 4.599 & 3.983 & 12.054 & 2134 & 2 & 3 & 50.2 & 32.24 & 0.1416 \\
    1.49 & $\mathrm{Put}^{2+}$ & 4.599 & 3.983 & 12.054 & 2134 & 2 & 4 & 50.2 & 32.24 & 0.1416 \\
    1.49 & $\mathrm{Spd}^{3+}$ & 4.599 & 3.983 & 12.054 & 2134 & 2 & 5 & 50.2 & 32.24 & 0.1416 \\
    1.49 & $\mathrm{Spm}^{4+}$ & 4.599 & 3.983 & 12.054 & 2134 & 2 & 6 & 50.2 & 32.24 & 0.1416 \\
    \midrule
    1.97 & $\mathrm{Eta}^{1+}$ & 5.695 & 4.932 &  9.838 & 3221 & 3 & 4 & 50.0 & 32.31 & 0.1405 \\
    1.97 & $\mathrm{Put}^{2+}$ & 5.695 & 4.932 &  9.838 & 3217 & 3 & 5 & 50.0 & 32.27 & 0.1405 \\
    1.97 & $\mathrm{Spd}^{3+}$ & 5.695 & 4.932 &  9.838 & 3212 & 3 & 6 & 50.0 & 32.22 & 0.1405 \\
    1.97 & $\mathrm{Spm}^{4+}$ & 5.695 & 4.932 &  9.838 & 3207 & 3 & 7 & 50.0 & 32.17 & 0.1405 \\
    \bottomrule
  \end{tabular}
  \end{table*}

Production trajectories ran for \SI{500}{ns} at a \SI{2}{fs} timestep, and coordinates were saved every \SI{20}{ps}, corresponding to $2.5\times10^8$ steps at a \SI{2}{fs} timestep. Equilibration was executed for \SI{10}{ns}, also at a \SI{2}{fs} timestep with thermostat coupling time \SI{0.5}{ps}. The number of replicates per system and field are specified in Table~\ref{tab:Sruns}. 

\begin{table}[tbp]
\centering\caption{Simulated trajectories per system and field. Each entry is a \SI{500}{\nano\second} trajectory. ``All'' denotes the 16 field strengths from $E=0$ to \SI{0.150}{\volt\per\nano\meter} in increments of \SI{0.015}{\volt\per\nano\meter} and 0.300, 0.450, 0.600, 0.750, and \SI{0.900}{\volt\per\nano\meter}.}
\label{tab:Sruns}
\small
\begin{tabular}{l l c c}
\toprule
Cell & Systems & Fields & $n$ at \SI{0.135}{V/nm} \\
\midrule
$R=0.68$ nm & parents, Spd variants & all & 3 \\
$R=1.02$ nm & parents & all & 3 \\
$R=1.49$, 1.97 nm & parents & all & 3 \\
$R=1.97$ nm, $2L_z$ & Eta, Spm & 0.135 & 3 \\
Free solution & parents & 0, 0.135 & 3 \\
              & parents & 0.3, 0.6, 0.9 & --- \\
\bottomrule
\end{tabular}
\end{table}

\section{Complete transport data}
\label{sec:transport}

\begin{figure*}[ht]
\centering
\includegraphics[width=\textwidth]{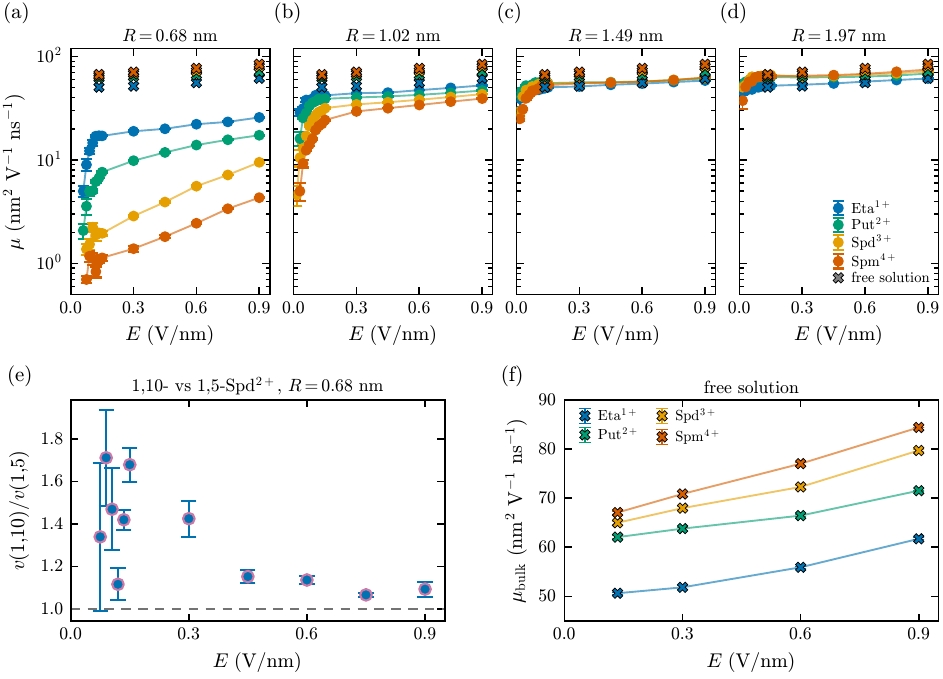}
\caption{Mobility across all simulated conditions. (a--d)~Mobility $\mu$ versus applied field $E$ for the four polyamines at $R=0.68$, 1.02, 1.49, and \SI{1.97}{\nano\meter}; crosses mark free-solution mobilities at the same fields. In the narrowest CNT the four species are ordered inversely to their charge. The spread between the amines is largest at low field; each mobility rises with field, and the multivalent polyamines' mobilities rise more steeply than $\mathrm{Eta}^{1+}$. At $R=\SI{1.02}{\nano\meter}$, the inversion is diminished but still present, and at $R\geq\SI{1.49}{\nano\meter}$, all amines approach their free-solution mobilities. (e)~Ratio $\frac{v_{1{,}10\text{-}\mathrm{Spd}^{2+}}}{v_{1{,}5\text{-}\mathrm{Spd}^{2+}}}$ at $R=\SI{0.68}{\nano\meter}$. The microspecies with separated protonation sites is faster at every resolved electric field strength by 7--71\%. (f)~Polyamine mobility in free solution versus applied field.}
\label{fig:S1}
\end{figure*}

Figure~\ref{fig:S1} presents the electrophoretic mobility $\mu=v/E$ of the four polyamines as a function of electric-field strength at each CNT radius ($R=\SI{0.68}{\nano\meter}$ data are shown in Figure~5a of the main text), together with the free-solution mobilities. Also presented is the field dependence of the velocity ratio between the two dicationic spermidine microspecies. 

\FloatBarrier
\section{Reproducibility}
\label{sec:reproducibility}
Three independent \SI{500}{\nano\second} simulation replicates were run at $E=\SI{0.135}{V/\nano\meter}$ for each parent amine at all radii and for each spermidine macrospecies at $R=\SI{0.68}{\nano\meter}$, all taken from a common equilibrium configuration and independent initial velocities (Table~\ref{tab:Spertraj} and Table~\ref{tab:Sreplicates}). The $\mathrm{Spd}^{3+}$ values are taken from the parent spermidine, while the last row of Table~\ref{tab:Sreplicates} is a second, independently built $\mathrm{Spd}^{3+}$ cell with identical composition. At other field strengths, results are taken from a single trajectory. The radius of gyration of spermidine varies by only 5\% across each of the varied protonation states, whereas that of the parent polyamines increases by a factor of 3.5 from $\mathrm{Eta}^{1+}$ to $\mathrm{Spm}^{4+}$. 

\begin{table}[ht]
    \centering
    \caption{Simulation replicate drift velocities, mobilities, and radii of gyration at $R=\SI{0.68}{\nano\meter}$. $n$ is the number of \SI{500}{\nano\second} trajectories executed; uncertainties are standard errors. $R_g$ denotes the radius of gyration of the solute which was sampled over \SI{1}{\nano\second} intervals beginning after the first \SI{50}{\nano\second} of production; $R_g$ via the same estimator is shown at every field in Figure~5c of the main text.}
    \label{tab:Sreplicates}
    \begin{tabular}{l c c r@{${}\pm{}$}l r@{${}\pm{}$}l c}
   System & $q$ & $n$ & \multicolumn{2}{c}{$v$ (nm/ns)}
      & \multicolumn{2}{c}{$\mu$ (nm$^2$ V$^{-1}$ ns$^{-1}$)} & $R_g$ (\AA) \\
    \midrule
    $\mathrm{Eta}^{1+}$ & $+1$ & 3 & 2.318 & 0.030 & 17.17 & 0.22 & 1.26 \\
    $\mathrm{Put}^{2+}$ & $+2$ & 3 & 0.908 & 0.004 &  6.72 & 0.03 & 2.12 \\
    $\mathrm{Spd}^{3+}$ & $+3$ & 3 & 0.266 & 0.007 &  1.97 & 0.05 & 3.29 \\
    $\mathrm{Spm}^{4+}$ & $+4$ & 3 & 0.143 & 0.008 &  1.06 & 0.06 & 4.43 \\
    \midrule
    $\mathrm{Spd}^{0}$                 & $0$  & 3 & 0.025 & 0.030 & \multicolumn{2}{c}{---} & 3.36 \\
    $1\text{-}\mathrm{Spd}^{1+}$      & $+1$ & 3 & 0.818 & 0.012 & 6.06 & 0.09 & 3.26 \\
    $1{,}5\text{-}\mathrm{Spd}^{2+}$  & $+2$ & 3 & 0.412 & 0.003 & 3.05 & 0.03 & 3.19 \\
    $1{,}10\text{-}\mathrm{Spd}^{2+}$ & $+2$ & 3 & 0.585 & 0.019 & 4.33 & 0.14 & 3.27 \\
    $\mathrm{Spd}^{3+}$  (independent build)  & $+3$ & 3 & 0.296 & 0.013 & 2.19 & 0.10 & 3.31 \\
    \bottomrule
    \end{tabular}
\end{table}

\FloatBarrier
\section{Construction and controls}
\label{si:construction}

\begin{table}[ht]
    \centering
    \caption{Drift velocity of every simulated trajectory at $E=\SI{0.135}{\volt\per\nano\meter}$ (nm/ns). The three replicates start from the same equilibrated configuration and each draws its own independent initial velocities from a random seed. Each velocity is a least-squares fit, discarding the first \SI{5}{ns}. The final column gives the mean velocity and standard error from the trajectories.}
    \label{tab:Spertraj}
    \small
    \begin{tabular}{l l c c c r@{${}\pm{}$}l}
    \toprule
    $R$ (nm) & System & 1 & 2 & 3 & \multicolumn{2}{c}{mean} \\
\midrule
0.68 & $\mathrm{Eta}^{1+}$ & 2.377 & 2.299 & 2.279 & 2.318 & 0.030 \\
 & $\mathrm{Put}^{2+}$ & 0.905 & 0.915 & 0.903 & 0.908 & 0.004 \\
 & $\mathrm{Spd}^{3+}$ & 0.275 & 0.270 & 0.253 & 0.266 & 0.007 \\
 & $\mathrm{Spm}^{4+}$ & 0.159 & 0.138 & 0.133 & 0.143 & 0.008 \\
 & $\mathrm{Spd}^{0}$ & 0.001 & $-0.011$ & 0.085 & 0.025 & 0.030 \\
 & $1\text{-}\mathrm{Spd}^{1+}$ & 0.829 & 0.831 & 0.793 & 0.818 & 0.012 \\
 & $1{,}5\text{-}\mathrm{Spd}^{2+}$ & 0.412 & 0.418 & 0.406 & 0.412 & 0.003 \\
 & $1{,}10\text{-}\mathrm{Spd}^{2+}$ & 0.582 & 0.620 & 0.554 & 0.585 & 0.019 \\
 & $\mathrm{Spd}^{3+}$ (independent build) & 0.270 & 0.313 & 0.305 & 0.296 & 0.013 \\
\midrule
1.02 & $\mathrm{Eta}^{1+}$ & 5.577 & 5.708 & 5.641 & 5.642 & 0.038 \\
 & $\mathrm{Put}^{2+}$ & 5.014 & 5.233 & 5.181 & 5.143 & 0.066 \\
 & $\mathrm{Spd}^{3+}$ & 4.016 & 4.076 & 4.088 & 4.060 & 0.022 \\
 & $\mathrm{Spm}^{4+}$ & 3.022 & 3.073 & 2.985 & 3.027 & 0.026 \\
\midrule
1.49 & $\mathrm{Eta}^{1+}$ & 6.717 & 6.692 & 6.640 & 6.683 & 0.023 \\
 & $\mathrm{Put}^{2+}$ & 7.468 & 7.593 & 7.647 & 7.569 & 0.053 \\
 & $\mathrm{Spd}^{3+}$ & 7.335 & 7.412 & 7.399 & 7.382 & 0.024 \\
 & $\mathrm{Spm}^{4+}$ & 7.027 & 7.014 & 6.934 & 6.992 & 0.029 \\
\midrule
1.97 & $\mathrm{Eta}^{1+}$ & 6.983 & 7.177 & 7.069 & 7.076 & 0.056 \\
 & $\mathrm{Put}^{2+}$ & 8.666 & 8.586 & 8.652 & 8.635 & 0.025 \\
 & $\mathrm{Spd}^{3+}$ & 8.803 & 8.847 & 8.827 & 8.826 & 0.013 \\
 & $\mathrm{Spm}^{4+}$ & 8.801 & 8.634 & 8.633 & 8.689 & 0.056 \\
\midrule
1.97, $2L_z$ & $\mathrm{Eta}^{1+}$ & 7.142 & 7.270 & 7.260 & 7.224 & 0.041 \\
 & $\mathrm{Spm}^{4+}$ & 8.539 & 8.856 & 8.530 & 8.642 & 0.107 \\
\midrule
free solution & $\mathrm{Eta}^{1+}$ & 6.906 & 6.801 & 6.816 & 6.841 & 0.033 \\
 & $\mathrm{Put}^{2+}$ & 8.426 & 8.308 & 8.403 & 8.379 & 0.036 \\
 & $\mathrm{Spd}^{3+}$ & 8.935 & 8.820 & 8.679 & 8.811 & 0.074 \\
 & $\mathrm{Spm}^{4+}$ & 9.130 & 9.038 & 8.996 & 9.055 & 0.040 \\
 \bottomrule
    \end{tabular}
\end{table}

Two controls for system construction were applied to every cell before use---firstly, carbons were binned by axial coordinate and minimum minimum-image carbon-to-carbon distance. An armchair CNT has an equal carbon count in every $z$-plane and a minimum separation equal to the carbon-carbon bond length. All production cells satisfy this with $d_{\mathrm{CC}}^{\min}=0.1405$--\SI{0.1416}{\nano\meter} (Table~\ref{tab:Sinventory}). Secondly, the water count was compared to that obtained when the same tube is equilibrated in contact with an orthorhombic water reservoir. Reservoir-equilibrated densities with end margins of 0.3--\SI{1.5}{\nano\meter} are  29.4, 31.0, 31.5, and \SI{32.1}{\per\nano\meter\cubed} for $R=0.68$, 1.02, 1.49, and \SI{1.97}{\nano\meter}, increasing toward the predicted bulk value with increasing radius (and, thereby, relaxed confinement).

At $R=\SI{0.68}{\nano\meter}$, the cells were solvated with \texttt{gmx solvate}. At $R=1.02$, 1.49, and \SI{1.97}{\nano\meter}, the solvation lattice results in a filling which deviates by several percent, so after \texttt{gmx solvate}, water counts were set to the target density taken from the reservoir equilibration for $R=\SI{1.02}{\nano\meter}$, \SI{32.25}{\per\nano\meter\cubed} at \SI{1.49}{\nano\meter}, and \SI{32.34}{\per\nano\meter\cubed} at \SI{1.97}{\nano\meter}. The deviations of the simulated cells from the respective reservoir-density values are $-1.9$ to $+2.4$\% at $R = 1.02$ and $\SI{1.49}{\nano\meter}$, $+0.2$ to
$+0.7$\% at $\SI{1.97}{\nano\meter}$, and $+0.6$ to $+2.3$\% at
$\SI{0.68}{\nano\meter}$ for Eta, Put and Spd. For spermine in the $R=\SI{0.68}{\nano\meter}$ CNT, the deviation is $-4.9$\% (\num{27.96} versus $\SI{29.4}{\per\nano\meter\cubed}$,  Fig.~\ref{fig:Sfilling}). Confined water counts reflect the solvation obtained under the protocol described. Table~\ref{fig:Sfilling} compares the water density in axial segments of the CNT interior more than \SI{1.2}{\nano\meter} from every solute atom with the same quantity in the reservoir-equilibrated cells. Every directly solvated cell lies within $+0.9$ to $+1.5$\% of the reservoir value. 

\begin{table}[ht]
\centering
\caption{Density of water in the $R=\SI{0.68}{\nano\meter}$ cells. $\rho$ is $N_{\mathrm{wat}}/V_{\mathrm{acc}}$ (Eq.~\ref{eq:salt}) and $\rho_{\mathrm{far}}$ is the density taken over \SI{0.35}{\nano\meter} axial increments from more than \SI{1.2}{\nano\meter} from a polyamine atom, averaged over 1000 frames of the equilibrium trajectory. The deviation is relative to $\rho_{\mathrm{far}}$ in reservoir-coupled solvation cells: 29.97 and \SI{30.04}{\per\nano\meter\cubed} for $\mathrm{Eta}^{1+}$ and $\mathrm{Spm}^{4+}$, and their mean for the other polyamines (\SI{30.01}{\per\nano\meter\cubed}). The cells were solvated with \texttt{gmx solvate}.}

\label{fig:Sfilling}
\begin{tabular}{l c c c c}
    \toprule
    System & $N_{\mathrm{wat}}$ & $\rho$ (nm$^{-3}$) & $\rho_{\mathrm{far}}$ (nm$^{-3}$) & Deviation (\%) \\
    \midrule
    $\mathrm{Eta}^{1+}$ & 944 & 29.59 & 29.70 & $-0.9$ \\
    $\mathrm{Put}^{2+}$ & 960 & 30.09 & 30.29 & $+0.9$ \\
    $\mathrm{Spd}^{3+}$ & 956 & 29.96 & 30.31 & $+1.0$ \\
    $\mathrm{Spm}^{4+}$ & 892 & 27.96 & 28.31 & $-5.8$ \\
    \midrule
    $\mathrm{Spd}^{0}$ & 963 & 30.18 & 30.46 & $+1.5$ \\
    $1\text{-}\mathrm{Spd}^{1+}$ & 962 & 30.15 & 30.46 & $+1.5$ \\
    $1{,}5\text{-}\mathrm{Spd}^{2+}$ & 958 & 30.03 & 30.31 & $+1.0$ \\
    $1{,}10\text{-}\mathrm{Spd}^{2+}$ & 958 & 30.03 & 30.35 & $+1.1$ \\
    $\mathrm{Spd}^{3+}$ (scaffold) & 956 & 29.96 & 30.27 & $+0.9$ \\
    \bottomrule
\end{tabular}
\end{table}

\section{Protocol Validation}
\label{sec:protocolvalidation}

The velocity-rescale thermostat\cite{bussi2007} acts on the entire system, including the drift of electrically driven solute. The electric field delivers power $P=\sum_i \mathbf{F}_i\cdot\mathbf{v}_i=E\sum_i q_i v_{z,i}$ to mobile atoms via Joule heating. In steady state the thermostat removes the Joule heat by uniformly rescaling velocities. $\mathrm{d}K/\mathrm{d}t=-2\gamma_{\mathrm{th}}K$ for uniform velocity damping $\gamma_{\mathrm{th}}$, so equating to $P$ gives $\gamma_{\mathrm{th}}=P/2K$, where $K$ is the kinetic energy of mobile atoms. This damping is negligible when compared with the rate $\frac{qeE}{mv}$ at which the solute's momentum relaxes. The ratio of the two rates, taken from the measured drift of solute and of both ionic species, is $7\times10^{-8}$ for $\mathrm{Eta}^{1+}$ and $2\times10^{-9}$ for $\mathrm{Spm}^{4+}$ at $R=\SI{0.68}{\nano\meter}$ and $E=\SI{0.135}{V/nm}$. The maximum value of this ratio---under any conditions---is $5.0\times10^{-5}$; therefore, direct thermostat forces on driven atoms do not affect the reported drift values.

\begin{figure*}
    \centering
    \includegraphics[width=\textwidth]{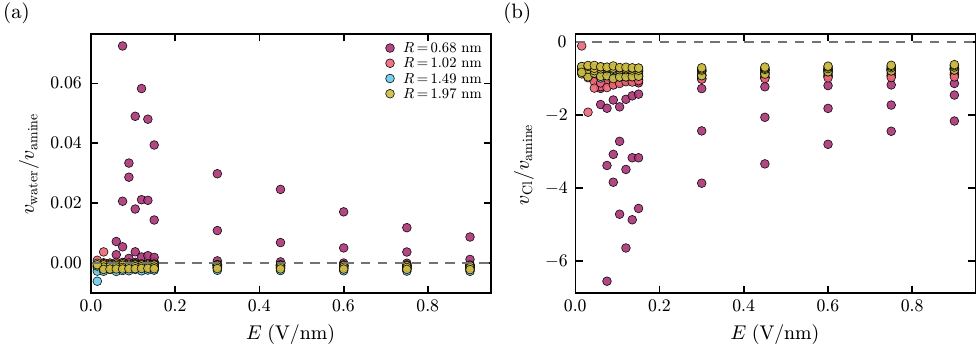}
    \caption{Water and ion drift velocities. (a)~The ratio of water drift velocity to amine drift velocity; remains uniformly small compared to that of the amines, showing that solvent advection does not account for the observed amine ordering. (b)~Drift ratio of chloride to amine. Chloride moves in the opposite direction as the cationic amines, ruling out co-advection as the cause of amine velocity.}
    \label{fig:S6}
\end{figure*}

\paragraph{Center-of-mass motion removal.} Linear center-of-mass (COM) motion was removed from the entire system every 100 timesteps. Since every cell is electroneutral, the applied field exerts no net force on mobile atoms; the removal then acts on fluctuations in total momentum. The measured drift of water remains below 8\% of solute drift under every condition (Figure~\ref{fig:S6}), which is consistent with this. 

\paragraph{Lateral periodic images.} Exterior to the CNT is vacuum, and neighboring periodic images can interact through PME. In order to bound this interaction, we increased the lateral cell dimensions of the $R=\SI{0.68}{\nano\meter}$ cell for $\mathrm{Eta}^{1+}$ and $\mathrm{Spm}^{4+}$, each by \SI{1.0}{\nano\meter}, from $2.720\times\SI{2.356}{\nano\meter}$ to $3.720\times\SI{3.356}{\nano\meter}$, leaving all other parameters unchanged (Table~\ref{tab:Sprotocol}). Both velocities are about 6\% higher in the expanded cell, and the ratio of their velocities remains constant.
\begin{table*}[ht]
    \centering
    \caption{Lateral cell-dimension sensitivity in the CNT of $R=\SI{0.68}{\nano\meter}$. The standard cell values are means taken from three \SI{500}{ns} trajectories (Table~\ref{tab:Sreplicates}), and the enlarged cell is a single \SI{200}{ns} trajectory with a five-block standard uncertainty.}
    \label{tab:Sprotocol}
    \begin{tabular}{l c c r@{${}\pm{}$}l r@{${}\pm{}$}l c}
    \toprule
    Cell & $L_x$ (nm) & $L_y$ (nm) & \multicolumn{2}{c}{$v_{\mathrm{Eta}}$ (nm/ns)} & \multicolumn{2}{c}{$v_{\mathrm{Spm}}$ (nm/ns)} & $v_{\mathrm{Eta}}/v_{\mathrm{Spm}}$ \\
    \midrule
    Standard & 2.720 & 2.356 & 2.318 & 0.030 & 0.143 & 0.008 & 16.2 \\
    Laterally extended & 3.720 & 3.356 & 2.449 & 0.023 & 0.152 & 0.008 & 16.2 \\
    \bottomrule
    \end{tabular}
\end{table*}

\FloatBarrier
\subsection{Axial cell length}
\label{si:length}

Because KCl concentration is approximately constant for each amine, radius and axial CNT length vary together (Table~\ref{tab:Sinventory}). To address this, we compared the $R=\SI{1.97}{\nano\meter}$ system at $L_z = 9.838$ and \SI{19.676}{\nano\meter}. By doubling the cell length, we double the accessible CNT volume, so we introduce additional KCl to reproduce the same \SI{50.0}{mM} KCl concentration. The total water count was set to roughly twice the reservoir-reference value. This then fixes radius, added KCl, and confined water density; however, only one amine is present in this test, so solute concentration is not matched. The velocities, though, differ by at most 2\% between the two cell lengths (Table~\ref{tab:Slength}).

\begin{table}[tbp]
\centering
\caption{Axial length control at $R=\SI{1.97}{\nano\meter}$ and $E=\SI{0.135}{\volt\per\nano\meter}$. Means and standard errors are reported over three replicates.}
\label{tab:Slength}
\begin{tabular}{l c r@{${}\pm{}$}l r@{${}\pm{}$}l}
\toprule
    $L_z$ (nm) & $N_{\mathrm{wat}}$
      & \multicolumn{2}{c}{$v_{\mathrm{Eta}}$ (nm/ns)}
      & \multicolumn{2}{c}{$v_{\mathrm{Spm}}$ (nm/ns)}
\\
\midrule
 9.838 & 3221 / 3207 & 7.076 & 0.056 & 8.689 & 0.056 \\
19.676 & 6429 / 6398 & 7.224 & 0.041 & 8.642 & 0.107 \\
\bottomrule
\end{tabular}
\end{table}

This comparison bounds the dependence of transport response on axial cell length, given that the control is performed on the shortest cell in the series.

\section{Free-solution comparisons}
\label{sec:controls}

To establish a baseline of unconfined transport, we perform NEMD simulations of each polyamine in free solution under the same force-field and thermostat conditions. Each system contains one polyamine in a \SI{5}{\nano\meter} cubic box of SPC/E water with ${\sim}\SI{50}{mM}$ KCl, each at both $E=0$ and \SI{0.135}{V\per\nano\meter} (three \SI{500}{\nano\second} trajectories each), and at 0.300, 0.600, and \SI{0.900}{V\per\nano\meter} (one trajectory each). The bulk ordering of mobility increases monotonically with valence, framing the observed inversion in $R=\SI{0.68}{\nano\meter}$ as a confinement effect (Table~\ref{tab:S2}).

\begin{table}[ht]
    \centering
    \caption{Free-solution drift velocities and mobilities for the four polyamines under four nonzero applied field strengths. At \SI{0.135}{V/nm}, values are means and standard errors from three \SI{500}{\nano\second} trajectories; higher fields use single trajectories with five-block uncertainties. Subscripts give the applied field in V/nm and mobilities are in nm$^2$\,V$^{-1}$\,ns$^{-1}$.}
    \label{tab:S2}
    \begin{tabular}{l c c c c c c}
    \toprule
    System & $q$ & $v_{0.135}$ (nm/ns) & $\mu_{0.135}$ & $\mu_{0.300}$ & $\mu_{0.600}$ & $\mu_{0.900}$ \\
    \midrule
    Ethylamine  & $+1$ & $6.84\pm0.03$ & $50.7\pm0.2$ & $51.9\pm0.3$ & $55.9\pm0.2$ & $61.7\pm0.2$ \\
    Putrescine  & $+2$ & $8.38\pm0.04$ & $62.1\pm0.3$ & $63.8\pm0.3$ & $66.4\pm0.2$ & $71.5\pm0.1$ \\
    Spermidine  & $+3$ & $8.81\pm0.07$ & $65.3\pm0.5$ & $68.0\pm0.3$ & $72.3\pm0.3$ & $79.7\pm0.2$ \\
    Spermine    & $+4$ & $9.06\pm0.04$ & $67.1\pm0.3$ & $70.8\pm0.4$ & $77.0\pm0.3$ & $84.4\pm0.1$ \\
\bottomrule
    \end{tabular}
\end{table}

\FloatBarrier
\section{Amine--counterion interactions and CNT wall force}
\label{sec:counterion}

\begin{figure*}[ht]
\centering
\includegraphics[width=\textwidth]{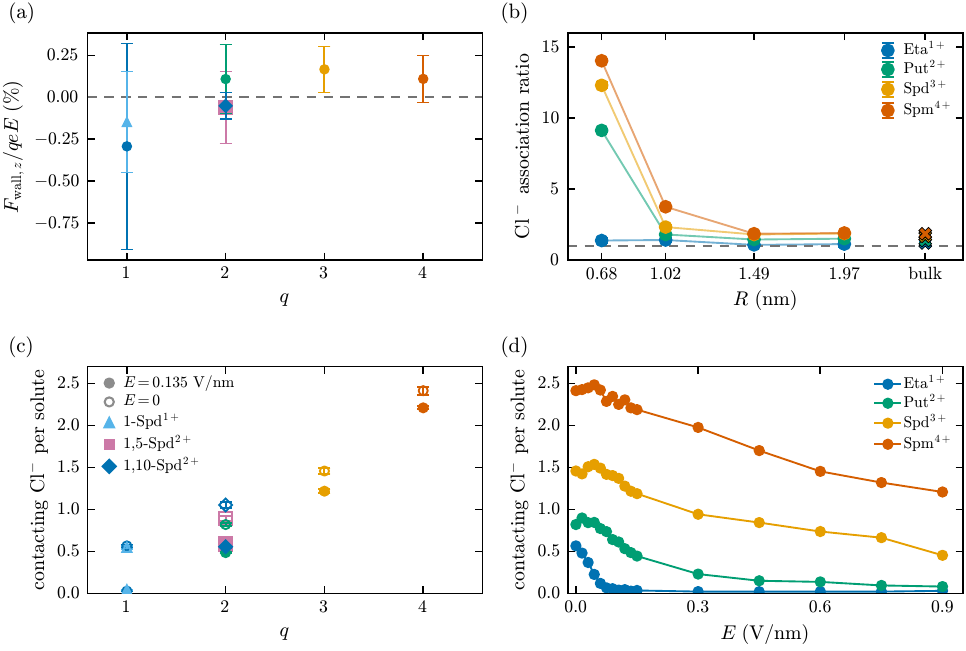}
\caption{Direct wall force and counterion contact in $R=\SI{0.68}{\nano\meter}$. (a)~Mean axial Lennard-Jones force exerted by the CNT on the amine, normalized by the electrical driving force at $E=\SI{0.135}{V/nm}$. Error bars are standard errors across the three trajectories. All values are consistent with zero. (b)~Chloride association ratio (relative to uniform mixing) versus CNT radius at $E=\SI{0.135}{V/nm}$; crosses are free solution, and the dashed line signifies that there is no association beyond that of mixing. (c)~Number of contacting chlorides per solute at $R=\SI{0.68}{\nano\meter}$; filled symbols are at $E=\SI{0.135}{V/nm}$ and open symbols $E=0$. Circles represent the fully protonated amines while the spermidine microspecies are identified by shape. The field dependence of this quantity is shown in Figure~5b of the main text.}
    \label{fig:Scounterion}
\end{figure*}

Figure~\ref{fig:Scounterion} characterizes the population of counterions around the solute and the direct force of the CNT on the solute. Each counterion was treated with a cutoff distance of $r_c=\SI{0.45}{\nano\meter}$ from any amine nitrogen. Sampling every \SI{0.5}{ns}, we discarded the first \SI{50}{ns} and estimated uncertainties from independent trajectories where they exist and five time blocks elsewhere. Cutoffs of 0.40 and \SI{0.50}{nm} gave the same qualitative trends. The direct wall force was computed as the axial component of the Lennard-Jones force exerted by all CNT carbons within the \SI{1.0}{\nano\meter} cutoff on every solute atom for every stored frame. The corresponding solute--CNT energies reproduce the energy-group values from GROMACS\cite{abraham2015} to within \SI{e-4}{kJ/mol}. Because the CNT carbons are uncharged, this is the complete wall contribution in the model. For each parent amine and spermidine microspecies at $R=\SI{0.68}{\nano\meter}$, the mean direct wall force is at most 0.3\% of $qeE$. The opposing force is therefore transmitted through the confined solvent rather than directly by the wall.

Because the chloride concentration differs between cells, the association ratio divides the contact count by that expected for chlorides distributed uniformly over the CNT,
\begin{equation}
    A = \frac{\langle n_c\rangle}{N_{\mathrm{Cl}}\,\langle V_c\rangle / V_{\mathrm{acc}}}.
    \label{eq:assoc}
\end{equation}

Here, $\langle n_c\rangle$ is the mean number of chlorides in contact with the species, $N_{\mathrm{Cl}}$ is the number of chlorides in the cell, and $V_c$ is the volume within $r_c$ of any solute nitrogen, all restricted to the radius that is accessible within the CNT, $R-\SI{0.17}{nm}$ (unrestricted in free solution) and estimated by Monte Carlo integration. $V_{\mathrm{acc}}$ follows from Eq.~\ref{eq:salt}, and in free solution is kept as the box volume. Therefore, $A=1$ denotes no association beyond uniform mixing. At $R=\SI{0.68}{\nano\meter}$ and $E=\SI{0.135}{V/nm}$, the polyamines carry 0.03, 0.49, 1.22, and 2.21 contacting chlorides in order of increasing charge, with association ratios of 1.4, 9.1, 12.3, and 14.0; for $R\geq\SI{1.49}{\nano\meter}$ and for free solution, this ratio is between 1.1 and 1.9.

If chloride association caused the inversion, the ordering should disappear when no chloride is near the solute. Therefore, we recomputed the drift velocity of each polyamine in the narrowest CNT using only frames when no chloride was within $r_c$ of any polyamine nitrogen. The inverted ordering persists without chloride contact. At $E=\SI{0.900}{V/nm}$, chloride-free $\mathrm{Eta}^{1+}$, $\mathrm{Put}^{2+}$, $\mathrm{Spd}^{3+}$, and $\mathrm{Spm}^{4+}$ migrate at 23.6, 16.0, 9.45, and \SI{3.08}{nm/ns}, respectively (with standard errors $\leq\SI{0.2}{nm/ns}$). The same ordering held at 0.450 and \SI{0.600}{V/nm}. 

\section{Static hydration structure}
\label{sec:structure}

Analysis of coordination numbers across the parent amines shows that terminal nitrogen sites have more similar coordination in the wider CNTs. For $R\geq\SI{1.02}{\nano\meter}$ at $E=\SI{0.135}{\volt\per\nano\meter}$, mean terminal-nitrogen coordination number ranges from 4.17--4.23 for ethylamine to 4.42--4.53 waters for spermine. In the $R=\SI{0.68}{nm}$ CNT, ethylamine is instead the most coordinated; however, hydration does not uniquely track the observed inversion in mobility---at $R=\SI{1.02}{\nano\meter}$, the ordering is inverted while the multivalent amines remain more coordinated to surrounding water than ethylamine. Consistent with the relaxation of local hydration by increasing CNT radius, though, this convergence in hydration accompanies the recovering of mobility ordering.

\begin{table}[tbp]
\centering
\caption{Water coordination per terminal nitrogen for each CNT radius at $E=\SI{0.135}{\volt\per\nano\meter}$, computed as the mean number of water oxygens within \SI{0.35}{\nano\meter} over terminal nitrogens. Values are the means $\pm$ standard errors from three independent trajectories sampled from 50--\SI{300}{ns}.}
\label{tab:parent_coordination}
\begin{tabular}{lcccc}
\toprule
& \multicolumn{4}{c}{CNT radius (nm)} \\
\cmidrule(lr){2-5}
Polyamine & 0.68 & 1.02 & 1.49 & 1.97 \\
\midrule
Ethylamine & $4.28\pm0.01$ & $4.17\pm0.02$ & $4.23\pm0.01$ & $4.20\pm0.01$ \\
Putrescine & $4.13\pm0.01$ & $4.22\pm0.01$ & $4.30\pm0.01$ & $4.27\pm0.01$ \\
Spermidine & $4.15\pm0.01$ & $4.39\pm0.03$ & $4.42\pm0.02$ & $4.38\pm0.01$ \\
Spermine & $4.06\pm0.01$ & $4.42\pm0.01$ & $4.53\pm0.01$ & $4.47\pm0.01$ \\
\bottomrule
\end{tabular}
\end{table}

\begin{figure*}
    \centering
    \includegraphics[width=\textwidth]{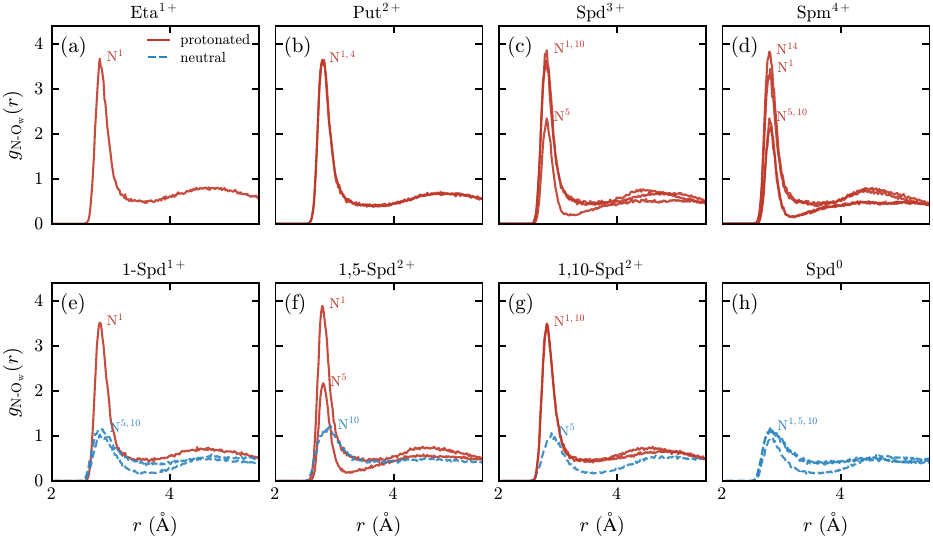}
    \caption{Per-nitrogen water radial density ratios between protonated and neutral nitrogens; i.e. $g_{\mathrm{N-O}_w}(r)$ in $R=\SI{0.68}{\nano\meter}$ at $E=\SI{0.135}{\volt\per\nano\meter}$. The top row contains the parent-polyamine data while the bottom row contains the spermidine-macrospecies data. In solid red are protonated N$^+$; in dashed blue are neutral N$^0$. Shell counts are normalized by the bulk water-oxygen density. Protonated nitrogens show a sharp first peak at $r \approx \SI{2.8}{\angstrom}$ with peaks of magnitude 2.2--3.9, while neutral nitrogens show a broader and weaker first peak of 1.0--1.3. The variation in peak heights between protonated nitrogens on the same molecule reflect differences in their local steric environments.}
    \label{fig:S7}
    \end{figure*}

\begin{figure*}[ht]
        \centering
        \includegraphics[width=\textwidth]{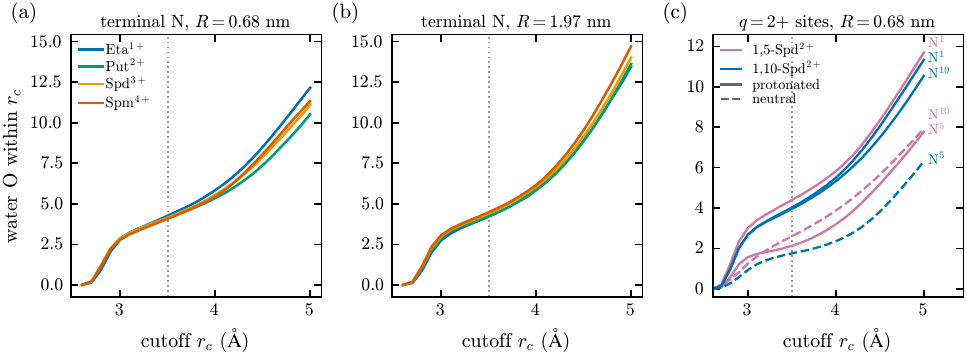}
        \caption{Water coordination number across each nitrogen site as a function of cutoff radius $r_c$; the dotted line marks the \SI{3.5}{\angstrom} cutoff used in the main text. (a,b)~Coordination for the terminal nitrogens of each of the four fully protonated polyamines at $E=\SI{0.135}{V/nm}$ in $R=0.68$ and \SI{1.97}{\nano\meter}, averaged over the trajectories in Table~\ref{tab:parent_coordination}. At $R=\SI{1.97}{\nano\meter}$, coordination scales approximately with charge. At $R=\SI{0.68}{\nano\meter}$, each species displays a comparable value for $r_c\leq\SI{3.5}{\angstrom}$, and at larger cutoff radii, ethylamine is the most coordinated. (c)~Coordination of $\mathrm{N}^1$, $\mathrm{N}^5$, and $\mathrm{N}^{10}$ for the two $q=+2$ spermidine microspecies at $E=\SI{0.135}{V/nm}$ (taken from three trajectories). Solid curves are the protonated nitrogens while dashed curves are the neutral ones. At every cutoff value, the coordination on the two protonated sites of $1{,}5\text{-}\mathrm{Spd}^{2+}$ differ far more than those of $1{,}10\text{-}\mathrm{Spd}^{2+}$ (4.39 and 2.13 compared to 4.03 and 3.96, respectively at \SI{3.5}{\angstrom}).} 
        \label{fig:S8}
    \end{figure*}

Figures~\ref{fig:S7} and \ref{fig:S8} characterize static hydration. We distinguish protonated and neutral nitrogen environments by radial density ratios; cumulative coordination curves quantify how both confinement and protonation topology alter the surrounding water environment. 

\clearpage
\bibliographystyle{unsrtnat}
\bibliography{references}